\documentclass[12pt]{article}
\usepackage{amsmath,amssymb}
\usepackage[margin=.8 in]{geometry}
\usepackage{hyperref}
\usepackage{mathtools}
\usepackage{color}
\usepackage{graphicx,float}
\usepackage{authblk}
\usepackage{commath}
\newcommand{\precc}{\mathrel{\prec\!\!\ast}}
\newcommand{\tprec}{\tilde{\prec}}
\newcommand{\tnprec}{\tilde{\nprec}}
\newcommand{\dd}{\text{d}}\newcommand{\ee}{\text{e}}\newcommand{\ii}{\text{i}}
\def\_#1{_{\scriptscriptstyle\text{#1}}^{~}}
\newcommand{\beq}{\begin{equation}}\newcommand{\eeq}{\end{equation}}
\newcommand{\bea}{\begin{eqnarray}}\newcommand{\eea}{\end{eqnarray}}
\begin{document}
\title{\bf A Tale of Two Actions \\ \large A Variational Principle for Two-Dimensional Causal Sets}
\author[1]{L. Bombelli\thanks{bombelli@olemiss.edu}}
\author[1]{B.B. Pilgrim\thanks{bbpilgri@go.olemiss.edu}}
\affil[1]{Department of Physics and Astronomy, University of Mississippi,
University, MS 38677-1848}
\date{October 23, 2020}
\maketitle
\vspace{10pt}
\abstract{\noindent In this paper we will explore two different proposals for the action for causal sets: the Benincasa-Dowker action \cite{Dowker}, and a modified version of the chain action \cite{chains}. We propose a variational principle for two-dimensional causal sets and use it for both actions to determine which causal sets at least on average satisfy a discrete version of the Einstein equation. Specifically, we test this method on causal sets embedded in 2d Minkowski, de Sitter, and anti-de Sitter spacetimes and compare these results to the most prominent nonmanifoldlike causal sets, Kleitman-Rothschild causal sets \cite{KR}.}

\section{Introduction}

Causal set theory (CST) is a theory of quantum gravity (for a full review, see Ref.\ \cite{review}) that replaces a Lorentzian manifold $(\mathcal{M},g_{\mu\nu})$ with a countable set of elements and a partial order $(\mathcal{C},\preceq )$. The partial order dictates the causal relations of the elements, i.e., if $i\preceq  j$ then $j$ is to the causal future of $i$. Much of the work done on causal sets is done in intervals: $I(p,q) = \{i\mid p\preceq  i \preceq  q\}$. Intervals are the causal set version of Alexandrov sets. In practice a sub partial order $\prec$ is used where $i\prec j\Leftrightarrow i\preceq  j$ and $j\neq i$. Some geometrical information can be directly extracted from the causal set: the volume of a region can be calculated by simply counting the number of elements in the region, and the timelike distance between two elements can be found by the length of the longest chain, $\{c_1,c_2,\cdots,c_{k}\mid c_1\prec c_2\prec \cdots\prec c_k\}$, between them. CST originated based on a Hawking-Malament theorem which states that in the continuum, a combination of the volume element and the structure of the light cones can be used to find every element of the metric \cite{HM-1,HM-2}. As we have both of these pieces of information, the number of elements in a causal set is a measure of discrete volume, and the partial order is seen as a discrete causal structure, we should be able to associate to a causal set all of the discrete geometrical information including the scalar curvature.

\section{CST Action} 

If we could find the scalar curvature at each element, we could recreate the Einstein-Hilbert action for general relativity, given by 
\beq
S\_{EH} = \int \sqrt{-g(x)}\, R(x)\, \dd^dx\;,
\eeq
where we set $c = 16\pi G_d = 1$. The causal set version of this will be something like
\beq
S\_{CST} =\ell^d\sum_{i\in\mathcal{C}} R_i\;,
\eeq
where $R_i$ is the scalar curvature at each element, and $\ell$ is a length-scale parameter which in the case of a causal set embedded in a manifold with density $\rho$ is taken to be the average continuum distance between the points, $\ell = \rho^{-1/d}$. We will restrict our attention to two dimensions and explore two proposals for this quantity.

\subsection{Benincasa-Dowker Action}

In Ref.\ \cite{Dowker}, Dionigi Benincasa and Fay Dowker proposed an action based on the d'Alembertian operator. Given some Cartesian coordinate system in two-dimensional Minkowski space, the massless Klein-Gordon equation reads
\beq
\Box \phi(x) = \bigg(-\frac{\partial^2}{\partial t^2}
+\frac{\partial^2}{\partial x^2}\bigg)\phi(x) = 0 \;.
\eeq
There's no trivial way to extend this to causal sets embedded in 2D Minkowski space; we only have a partial order and no coordinates, and we can't literally take derivatives. Additionally, because of the nature of causal sets, any operator we use to replace the d'Alembertian must be nonlocal; furthermore, we should require it to be retarded (i.e., the field at any point should only depend on the field on the past light cone), linear, and Lorentz invariant.\footnote{The causal set version of Lorentz invariance (or general covariance) is label invariance. All physical quantities should be independent of any label assigned to each element.} One possibility suggested by Sorkin in Ref.\ \cite{Sork} is to sum over layers in the causal set. A $k$-layer $L_k$ relative to some $i \in \mathcal{C}$ consists of the elements $j$ such that $I(j,i)$ contains $k+1$ elements. The causal set d'Alembertian then is
\beq
B\phi_i = \frac{1}{\ell^2}\bigg(a_0\phi_i+\sum_{k=1}^\infty a_k \sum_{j\in L_k} \phi_j\bigg),
\label{Bphii}
\eeq
where the coefficients are found by requiring that for causal sets embedded with increasing density the expected value of $B\phi_i$ tends to the continuum value,
\beq
\lim_{\ell\to 0}\left<B\phi_i\right> = \Box\phi(x_i) \;. \label{box}
\eeq
This doesn't define a single d'Alembertian but rather a family of them the simplest of which is found by setting $a_0 = -2, a_1 = 4, a_2 = -8, a_3 = 4,$ and $a_k = 0$ $\forall k\geq 5.$ We can find the expected value in Eq.\ (\ref{box}) for causal sets embedded in Minkowski space by assuming the elements of the causal set are distributed in the manifold via a Poisson distribution. According to this distribution, the probability of finding $n$ points in a region of volume $V$ is
\beq
P(n,V) = \ee^{-\rho V}\frac{\left(\rho V\right)^n}{n!} \;,
\eeq
where $\rho = N/V = 1/\ell^2$ is the density. The $a_1^{~}$ term can then be obtained in this setting by identifying $\sum_{j\in L_1}\phi_j^{~}$ with
\beq
\int_{J^-(x_i)}\ee^{-\rho V(x_i,y)}\phi(y)\,\dd^2y \;.
\eeq
At each point of integration $y$, we have the probability that there are no points between the evaluation point $x_i$ and $y$ multiplied by the field at that point. This is the continuum equivalent of summing over all values of the field at elements one layer from the evaluation element. We can do something similar for the other terms:
\beq
\left<B\phi_i\right> = \frac{1}{\ell^2}\left[-2\phi(x_i)+\frac{1}{\ell^2}\int_{J^-(x_i)}\dd^2y\ \ee^{-\rho V(x_i,y)}\left(4-8\rho V(x_i,y)+2(\rho V(x_i,y))^2\right)\phi(y)\right].
\label{Bphi}
\eeq
To relate this expression to the action, we consider what happens to it in curved spacetime. If we wanted to, we could evaluate it for specific spacetimes, but to be general, we can expand the metric in Riemann normal coordinates \cite{gravity}. To first order in $\mathcal{R}L^2$ where $\mathcal{R}$ is any component of the Riemann tensor, Ricci tensor, or scalar curvature, and $L$ is the size of the Riemann normal coordinate neighborhood we're considering,
\beq
g_{\mu\nu}(x) = \eta_{\mu\nu}-\textstyle{\frac13}\,(x-x_0)^\alpha (x-x_0)^\beta
R_{\mu\alpha\nu\beta}(x_0^{~}) \;,
\eeq
and
\beq
\sqrt{-g(x)} = 1-\textstyle{\frac16}\,(x-x_0)^\alpha(x-x_0)^\beta R_{\alpha\beta}(x_0^{~}) \;.
\eeq
The volume element then becomes $\dd^2y \to \sqrt{-g(y)}\,\dd^2y$, and $V(x_i,y) = \int_{J^-(x_i)\cap J^+(y)}\dd^2z \to \int_{J^-(x_i)\cap J^+(y)}\sqrt{-g(z)}\,\dd^2z$. Calculating this integral and using it in Eq.\ (\ref{Bphi}) yields
\beq
\lim_{\ell\to 0}\left<B\phi_i\right>
= \left(\Box-\textstyle{\frac12}R(x_0)\right)\phi(x_i) \;.
\eeq
This tells us that in the continuum one can obtain the scalar curvature by setting $\phi(x) = -2$ in a neighborhood of $x$, and we can then use Eq.\ (\ref{Bphii}) to calculate the action by setting $\phi_i = -2$ for all $i$ and summing over all elements of the causal set,\footnote{We add a multiplicative constant $2$ as we're calling $\int R(x)\text{d}^dx$ the action instead of $\frac{1}{2}\int R(x)\text{d}^dx$.}
\beq
S\_{BD} = 2\,(N-2N_1+4N_2-2N_3) \;,
\eeq
where $N$ is the number of elements, and $N_i$ is the number of intervals in the causal set which contain $i+1$ elements including the end points.

\subsection{Chain Action}

Alternatively, one can find the scalar curvature of a causal set from the number of chains. In Ref.\ \cite{Meyer}, Meyer calculated the expected number of chains for causal sets embedded in Minkowski space. We can derive this result by using the Poisson distribution: the probability that an infinitesimal volume contains one point is $\rho\dd^dx+\mathcal{O}((\rho\dd^dx)^2)$. If we consider an Alexandrov set where the minimal and maximal points are elements of the embedded causal set, any totally ordered $k$-element subset will make a chain of length $k+1$ between the maximal and minimal element.\footnote{As a reminder, our notation is different from Meyer's.} Thus, the probability that there's a chain of length $k+1$ through infinitesimal volumes $\dd^dx_i^{~}$ such that $\dd^dx_i^{~}\in J^+(\dd^dx_{i-1}^{~})\ \forall i$ is
\beq
\rho\,\dd^dx_1^{~}\,\rho\,\dd^dx_2^{~}\cdots\rho\,\dd^dx_k^{~}\;.
\eeq
The total expected number of chains of length $k$ then is
\beq
\left<c_k\right> = \rho^{k-1}\int_{A_0}\dd^dx_1\int_{A_1}
\dd^dx_2\cdots\int_{A_{k-2}}\dd^dx_{k-1} \;,
\eeq
where $A_i =J^+(x_i)\cap J^-(x_{k})$, $x_0^{~}$ is the minimal element, and $x_{k}^{~}$ is the maximal element. This integral can be calculated
\beq
\left<c_k\right>
= \frac{N^{k-1}}{k-1}\left(\frac{\Gamma(d+1)}{2}\right)^{k-2}
\frac{\Gamma(d/2)\,\Gamma(d)}{\Gamma((k-1)d/2)\,\Gamma(kd/2)} \;,
\eeq
where once again $N$ is the total number of points in the interval. As with the BD action, following Ref.\ \cite{chains} this can be generalized to an arbitrary spacetime by using Riemann normal coordinates and $\mathcal{O}((\mathcal{R}L^2)^2)$,
\beq
\left<c_k\right>_{\text{curved}} = \left<c_k\right>_{\text{flat}}
\left[1-\left(\frac{N}{\rho}\right)^{\!2/d} \frac{d\,(k-1)A_d}{12\,(dk+2)}
\left(\frac{1}{d\,(k-1)+2}\,R(x_0^{~}) - R_{00}^{~}(x_0^{~})\right)\right],
\eeq
where
\beq
A_d = \left(\frac{2^{d-2}\,d(d-1)\,\Gamma((d-1)/2)}{\pi^{(d-1)/2}}\right)^{\!2/d}
\eeq
and $R_{00}^{~}(x_0^{~})$ is the $00$ component of the Ricci tensor. Note we wish to recreate the Einstein-Hilbert action, $\int R(x)\sqrt{-g(x)}\dd^dx$, which in this regime where the curvature is approximately constant is equivalent to $R(x_0)V = R(x_0)N/\rho$. To find this quantity for a particular causal set, we can exchange the expected number of chains of length $k$ with the actual number and solve for this. To eliminate $R_{00}^{~}(x_0^{~})$, we'll need to choose 2 values of $k$, $k_1$ and $k_2.$ The simplest choice is to use the two lowest possible values, $k_1 = 3$ and $k_2=4$.\footnote{There are always exactly one chain of length one and $N$ chains of length two, so $c_1$ and $c_2$ carry no curvature information.} We will also restrict our attention to two dimensions. In this case,
\beq
S = \frac{8640}{N^3}\,c_4^{~}-\frac{1152}{N^2}\,c_3^{~}+48 \;.
\eeq

\subsection{Modification of the Chain Action}

In Ref.\ \cite{us} we proposed a modification of the expected number of chains which we will briefly describe here. So far both of the methods of obtaining the action we've explored have relied on the Poisson distribution; is this the correct distribution to use? It depends on the application we want to use the action for. As of now causal sets is really a framework of a discrete theory of classical gravity; to quantize it, we will have to use the path integral approach as canonical quantization makes little sense in this context. Given some appropriate boundary conditions,\footnote{What exactly is a boundary condition in causal set theory? It's unclear; in general relativity one uses the spatial metric and its derivatives on an initial and final spacelike hypersurface. The causal set equivalent would be a pair of antichains, a group of mutually unrelated points; however, antichains contain no information other than their cardinality. Perhaps a pair of thickened antichains would work.} the probability amplitude of evolving from one state to another is
\beq
\sum_{\mathcal{C}\,\in\,{\bf C}} \ee^{\ii S[\mathcal{C}]} \;,
\eeq
where ${\bf C}$ is some set of causal sets specified for the problem. For the action formulation of a classical field other than gravity, we would specify some coordinate system and the configuration of the field on an initial and final hypersurface and keep the time between these hypesurfaces as measured in the coordinate system constant (this is well-known, but see Ref.\ \cite{wald} for instance.) In general relativity, as the metric itself is the variable, we can't hold the time interval constant. To fix this, one proposal used in other theories of quantum gravity as well as classical general relativity is to hold the spacetime volume constant \cite{volume,volume2}. The causal set equivalent of this is holding the total number of points constant:
\beq
\sum_{\mathcal{C}\,\in\,{\bf C}_N} \ee^{\ii S[\mathcal{C}]} \;,
\eeq
where ${\bf C}_N$ is some subset of the set of all $N$ element causal sets. This is irreconcilable with the Poisson distribution. Using the Poisson distribution, a region of volume $V$ which on average has some density $\rho$ has a nonzero probability to have any number of points within it from $0$ to $\infty$. The binomial distribution, however, is well suited for this; the binomial distribution dictates that some region of volume $V_0$ has exactly $N$ points within it; thus, if some subset of the volume, $V,$ contains $k$ points, the rest of the volume $V_0-V$ must contain $N-k$ points. As such, the probability can be written as 
\beq
{P}(n,V) = {N \choose {n}}\left(\frac{V}{V_0}\right)^n
\left(1-\frac{V}{V_0}\right)^{N-n}.
\eeq
We can now recalculate the chain length distribution for causal sets embedded in $d$-dimensional Minkowski space; the probability that there's one point in some differential volume $\dd^dx$ (and the rest of the points can be anywhere in $V_0$) is $N\dd^dx/V_0 = \rho\, \dd^dx$. The deviation begins now; given that there is already a point in a differential volume, the probability that there is one point in another differential volume is $(N-1)\, \dd^dx'/V_0$. The fact that one of the $N$ points has already been allocated, alters the probability for the next point; this happens repeatedly. The expected number of chains of length $k$ can then be calculated as
\beq
\left<c_k\right>
= \rho_0\int_{A_0}\dd^dx_1\rho_1\int_{A_1}\dd^dx_2
\cdots\rho_{k-2}\int_{A_{k-2}}\dd^dx_{k-1} \;,
\eeq
where $\rho_i = (N-i)/V_0$. This is the same integral as before albeit with a different coefficient, so it can be evaluated:
\begin{equation}
\left<c_k\right> = \frac{N!}{(N-(k-1))!\,(k-1)}\left(\frac{\Gamma(d+1)}{2}\right)^{k-2}\frac{\Gamma(d/2)\,\Gamma(d)}{\Gamma((k-1)d/2)\,\Gamma(kd/2)}\;.
\end{equation}
This equation says that the number of chains of length greater than $N+1$ is identically $0$ as one would expect; similarly, there are exactly $N$ chains of length two. Expanding in Riemann normal coordinates and restricting ourselves to two dimensions, we find
\beq
\left<c_k\right> = \frac{N!}{(N-(k-1))!}\left[\frac{1}{((k-1)!)^2}+\frac{NR}{\rho}\left(\frac{2k^3-3k^2+k}{12\,(k!)^2}-\frac{k-1}{8\,((k-1)!)^2}\right)\right],
\eeq
where we have also used the fact that in two dimensions all components of the Riemann tensor can be expressed in terms of the scalar curvature. This allows us to find the action from the number of chains of any single length $k$. Once again, this isn't a single action but a family of them:
\beq
S_k = \left(\frac{2k^3-3k^2+k}{12\,(k!)^2}-\frac{k-1}{8\,((k-1)!)^2}\right)^{-1} \left(\frac{(N-(k-1))!}{N!}c_k-\frac{1}{((k-1)!)^2}\right).
\eeq
The simplest case is $k=3$ which we will denote $S\_{C}$ (for chains) and is given by
\beq
S\_{C} = 36\left(\frac{4c_3}{N(N-1)}-1\right).
\eeq
It might seem as though we should modify the BD action as well as it was calculated using the Poisson distribution; however, the coefficients were chosen in the infinite density limit where the Poisson and binomial distributions coincide. It's these two actions, $S\_{BD}$ and $S\_{C}$ which we will explore.

\section{Accuracy of the Actions}

As we have now introduced both actions, we wish to test them to see how closely they reproduce the continuum action. The only way we have of generating manifoldlike causal sets, causal sets which can be embedded in a manifold with an approximately constant density (and no continuum length scales shorter than average distance between the points), is by sprinkling. Sprinkling is a process by which we take a continuum Alexandrov set in a Lorentzian manifold, and choose points within it uniformly at random to. A causal set is obtained by inducing a partial order among those points determined by the continuum causal structure; in other words $i\prec j$ if and only if the locations $x_i$ and $x_j$ of elements $i$ and $j$, satisfy $x_j\in I^+(x_i)$.

\subsection{The Mechanics of Sprinkling}

\begin{figure}
\begin{center}
\includegraphics[width=.45\textwidth]{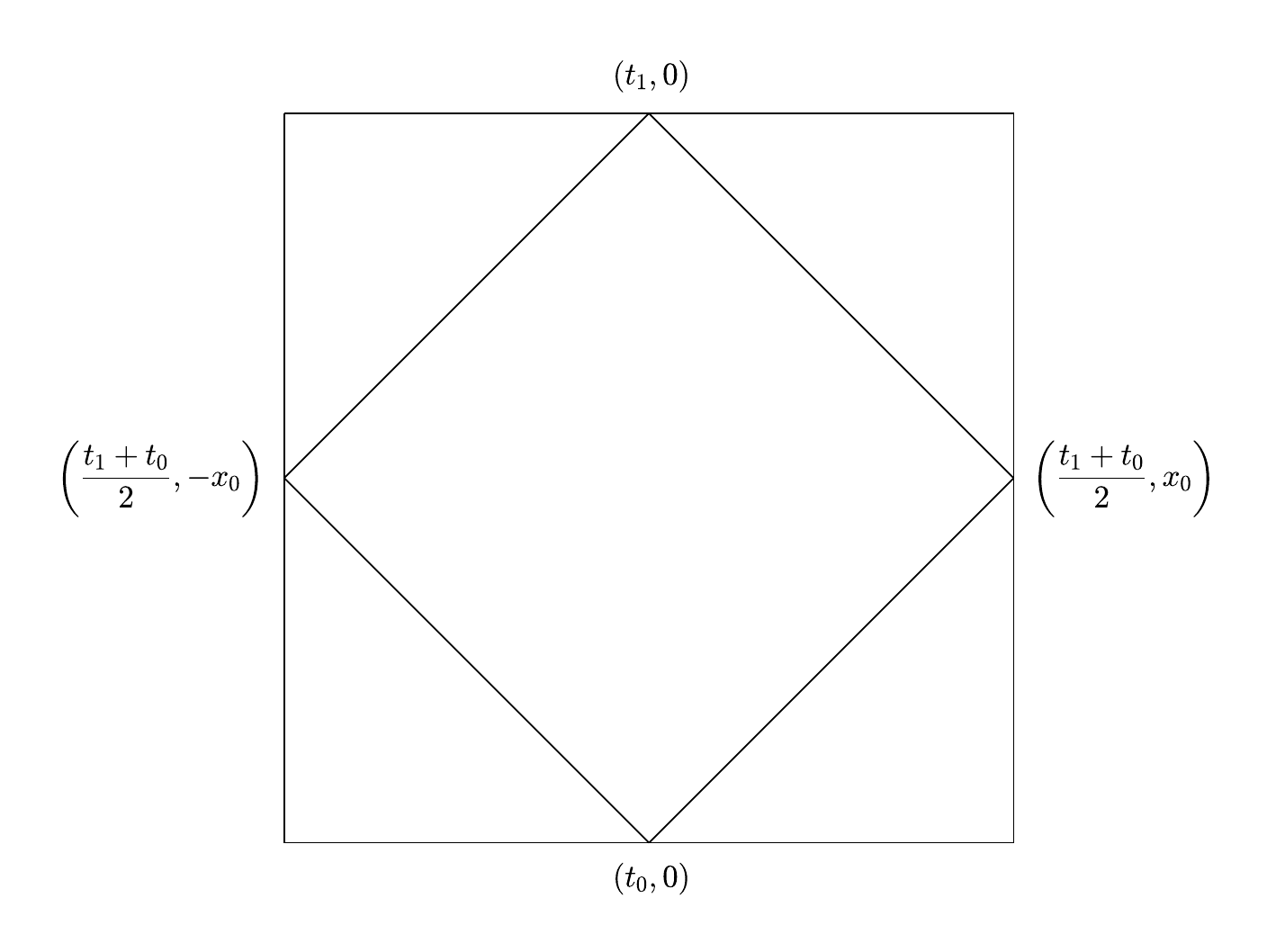}
\end{center}
\caption{An Alexandrov set in two-dimensional Minkowski space (or any spacetime in conformal coordinates.) To sprinkle, we put points in the square around it and only keep the ones that fall inside the diamond.}
\label{SA}
\end{figure}

The mechanics of sprinkling are fairly simple at least in Minkowski space. In some Cartesian coordinate system on two-dimensional Minkowski space, we choose an Alexandrov set such that the maximal and minimal points are along the direction of time.\footnote{This is arbitrary; we only make this choice because it makes figures look nice.} In Fig.\ \ref{SA}, we show such an Alexandrov set and the square that surrounds it; to sprinkle points within the set, we randomly select a value between $t_0^{~}$ and $t_1^{~}$ and another between $-x_0^{~}$ and $x_0^{~}$. We let the first value be the time coordinate of some point and the second its spatial coordinate. If this point lies in the Alexandrov set, we keep it and choose it to be the location of an element of the causal set. If it doesn't we discard it and choose another point. We do this repeatedly until we have the desired number of elements.

Sprinkling points into other spacetimes is more complicated. Although there has been some success in sprinkling in Schwarzschild spacetime \cite{BH}, the computational requirements are much lower to sprinkle in conformally flat spacetimes.\footnote{Luckily, all two-dimensional spacetimes are locally conformally flat.} We'll restrict our attention to these; specifically, we'll look at de Sitter and anti-de Sitter spacetimes. For de Sitter we'll use coordinates in which the line element reads
\beq
\dd s^2 = \frac{1}{H^2t^2}\left(-\dd t^2+\dd x^2\right) ,
\eeq
where $t\in(-\infty,0)$ and $x\in(-\infty,\infty)$. In this case, the volume element is $\sqrt{-g(x)}\, \dd t\, \dd x = (H^2t^2)^{-1} \dd t\,\dd x$; because it's not independent of $t$, we can't simply pick a random $t$ value between $t_1$ and $t_0$ as $\sqrt{-g(x)}$ increases with $t$, and in these coordinates the apparent density should be higher in the top part of the Alexandrov set.\footnote{We say apparent density because the actual density is still approximately constant, but it doesn't appear to be the case because our minds are wired to think in terms of Euclidean space.} To account for this, we can define the following quantity
\beq
r = \frac{\displaystyle\int_{t_0}^{\bar{t}}\frac{\dd t}{H^2t^2}}{\displaystyle\int_{t_0}^{t_1}\frac{\dd t}{H^2t^2}} \;, \label{r}
\eeq
where $r\in[0,1]$, and $\bar{t}\in[t_0,t_1]$, which represents the fraction of the total timelike distance between $t_0$ and $t_1$ up to $\bar{t}$. If we generate a uniformly random $r$ the corresponding $\bar{t}(r)$,
\beq
\bar{t} = \frac{t_0\,t_1}{r(t_0-t_1)+t_1} \;,
\eeq
obtained by inverting Eq.\ (\ref{r}), is distributed according to the volume element. As the volume element is independent of $x$, $x$ can be chosen the same way as in Minkowski space. This will once again sprinkle points in the square around the Alexandrov set, and those not in the diamond can be excluded. Anti-de Sitter works the same way but with the line element
\beq
\dd s^2 = \frac{1}{H^2x^2}\left(-\dd t^2+\dd x^2\right)
\eeq
where $t\in(-\infty,\infty)$ and $x\in(0,\infty)$.
\begin{figure}[H]
\begin{center}
\includegraphics[scale = .5]{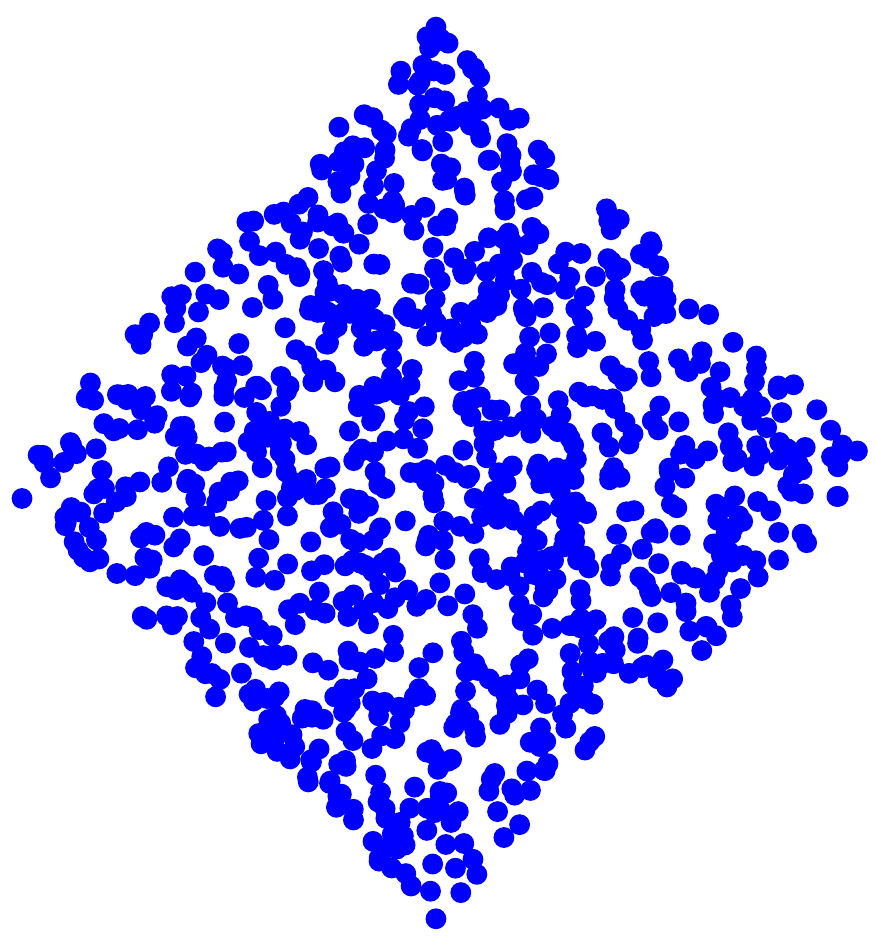}
\\
\includegraphics[scale = .5]{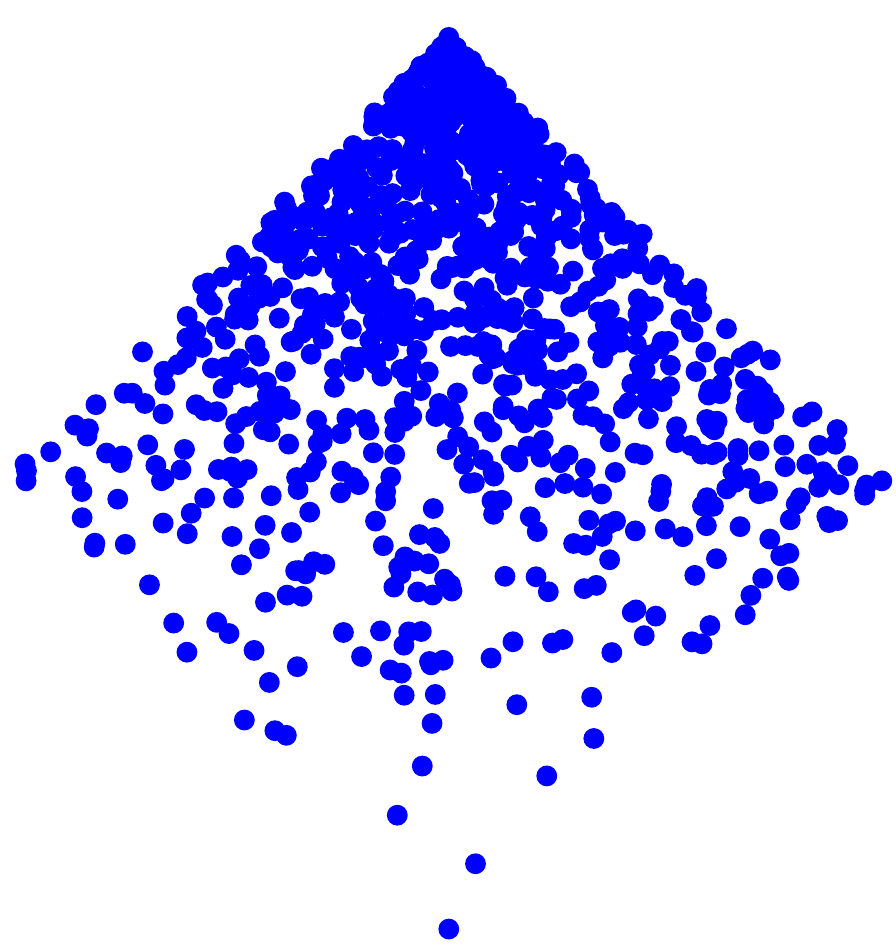}
\includegraphics[scale = .5]{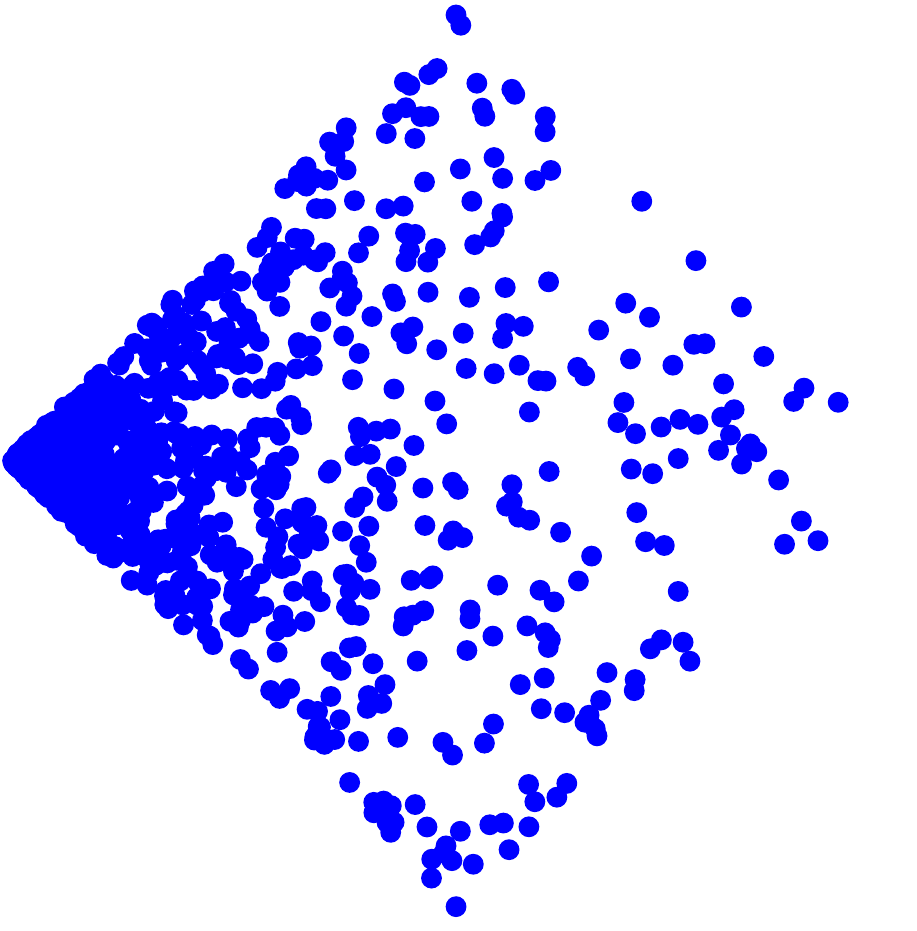}
\end{center}
\caption{$1000$-element causal sets plotted in Minkowski (top), de Sitter (bottom left), and anti-de Sitter space (bottom right). In both the de Sitter and anti-de Sitter cases, $H = 100$. Each blue dot represents an element of the causal set.}
\label{csets}
\end{figure}

\subsection{Results}

To calculate the actions, we sprinkle sets of points of various cardinalities in these spaces, and find the relations matrix, defined as
\beq
R_{i,j} =\begin{cases}
1 \hspace{1cm}\text{if $i\prec j$}\\
0 \hspace{1cm}\text{else} \end{cases}.
\eeq
From this matrix, we can find all of the quantities we need. For the chain action, the number of three-chains between the minimal and maximal points can be found via
\beq
c_3 = R^3_{0,N+1} = \sum_{i,j} R_{0,i}^{~}\, R_{i,j}^{~}\, R_{j,N+1}^{~}
\eeq
where $0$ is the minimal point and $N+1$ is the maximal point. In this notation, there are $N$ sprinkled points, and $N+2$ total points including the maximal and minimal ones. To find the BD action, we'll need the total number of $k$-element intervals. Note that $R^2_{i,j} \equiv (R^2)_{i,j}^{~}$ is the number of two-chains between $i$ and $j$ or equivalently the cardinality of $I(i,j)$. Thus,
\beq
N_{k+1} = \sum_{i,j} \delta(1,R_{i,j})\, \delta(k,R^2_{i,j})\;,
\eeq
where $\delta(l,m)$ is the Kronecker delta and $\delta(1,R_{i,j})$ is needed because intervals are only formed by related points; this is only a problem for $k=0$ because $R^2_{i,j} = 0$ for all points such that $i\nprec j$ yet these do not contribute to $N_1$.

We can now calculate the actions. Specifically, for each type of spacetime and for $N=100$ to $N=2000$, we sprinkle 100 causal sets, calculate the action, and average it.
\begin{figure}
\begin{center}
\includegraphics[scale = .5]{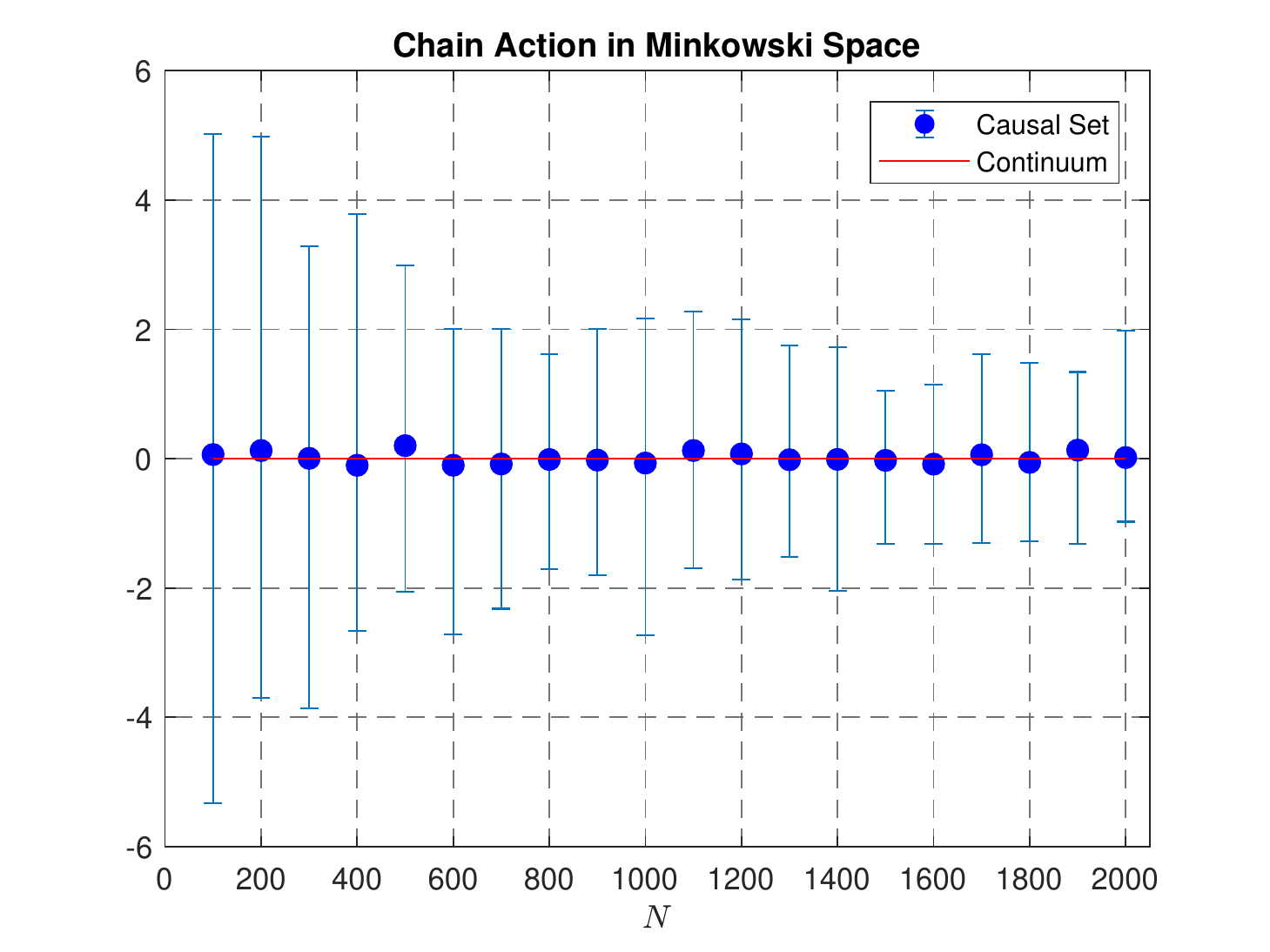}
\includegraphics[scale = .5]{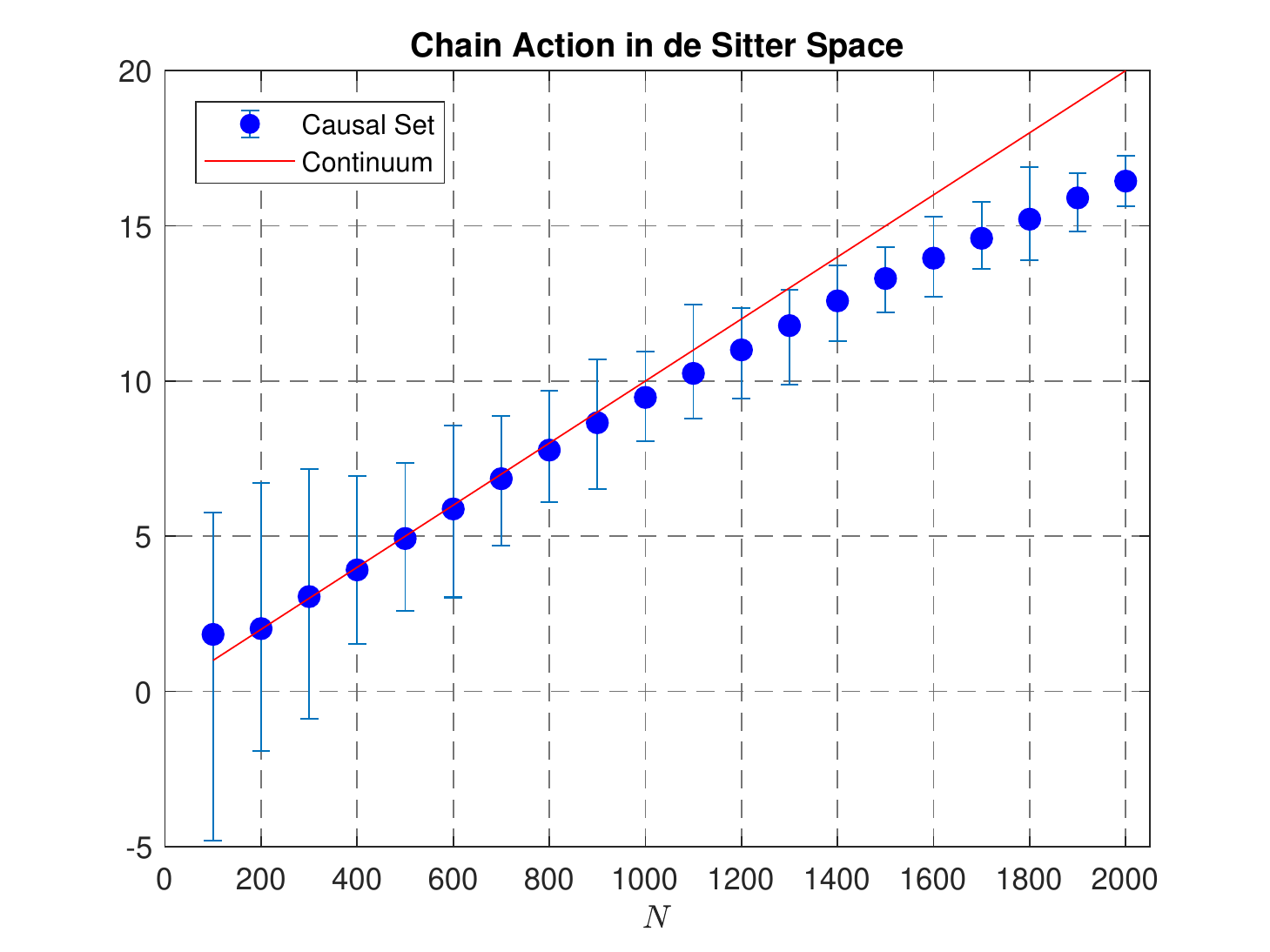}
\includegraphics[scale = .5]{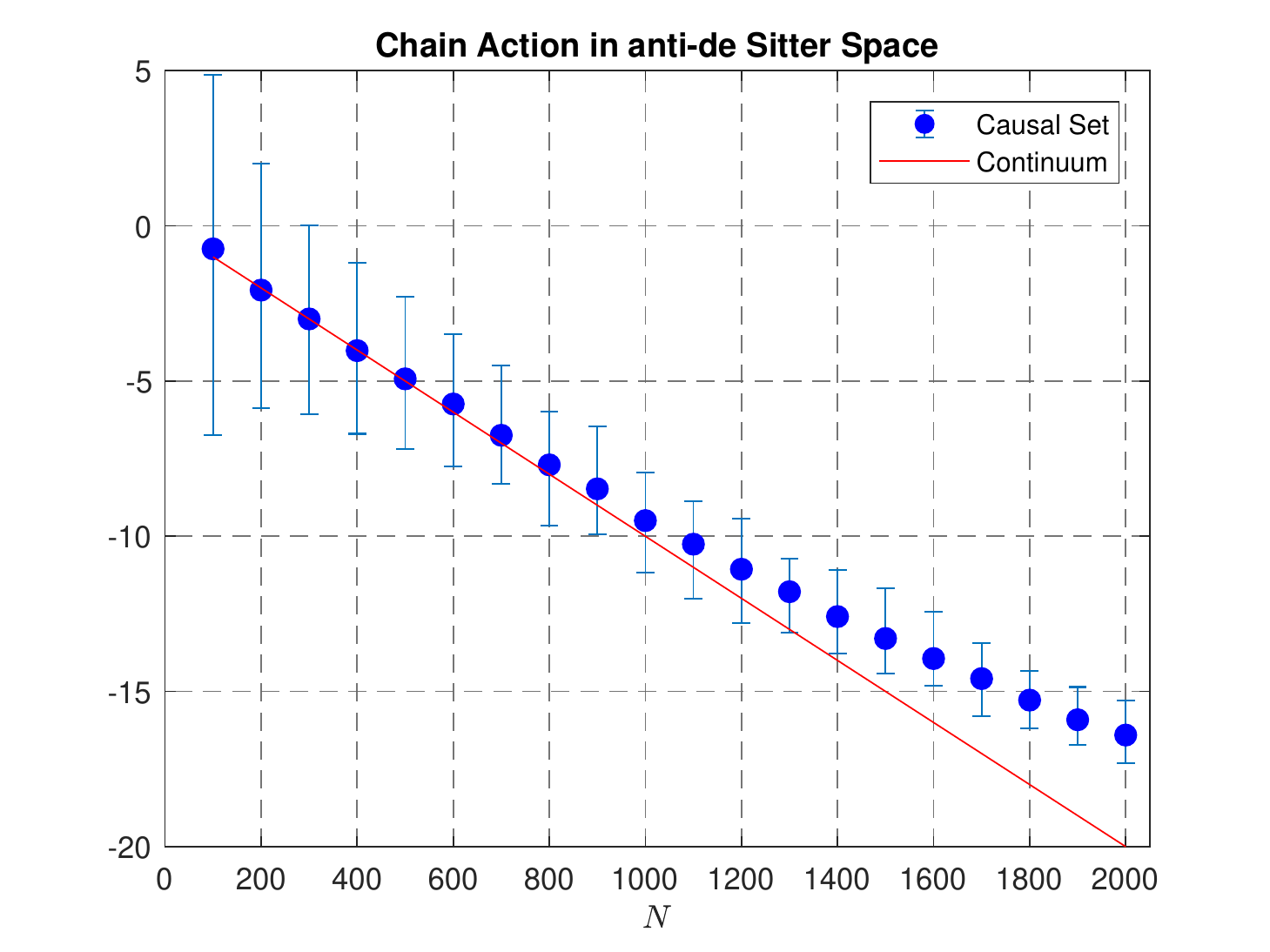}
\end{center}
\caption{Plots of the chain action vs.\ $N$ for Minkowski, de Sitter, and anti-de Sitter spaces. For both de Sitter and anti-de Sitter space, $H = 100$ and the point density $\rho = 2\times 10^6$ are constant; thus, as $N$ grows the size of the Alexandrov set grows compared to the curvature scale. For each size $N$ we show the average value of the action $S\_C$ over 100 different causal sets, and the error bars range from the causal set with the highest action to the causal set with the lowest action.}
\label{cact}
\end{figure}
\begin{figure}
\begin{center}
\includegraphics[scale = .5]{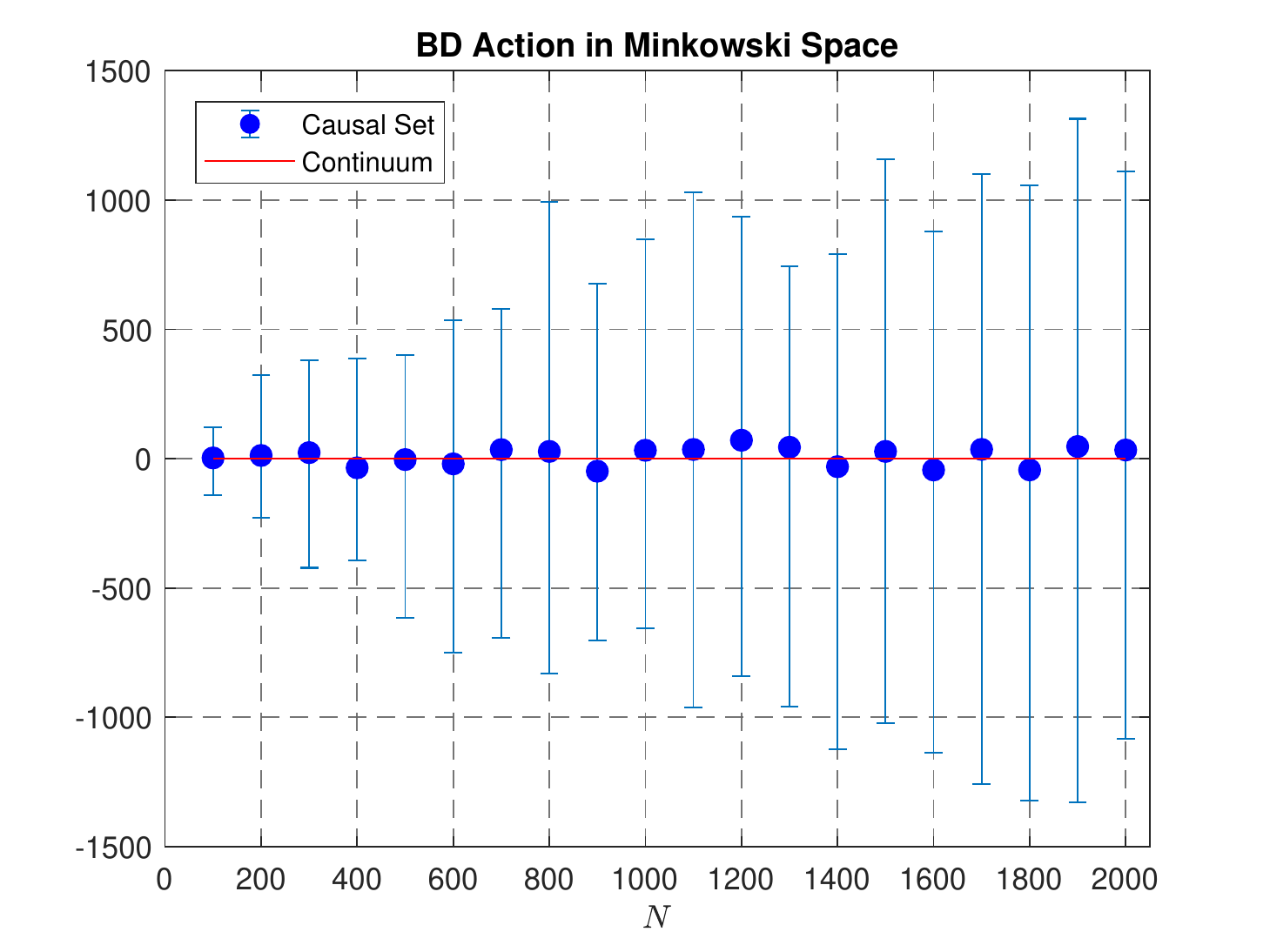}
\includegraphics[scale = .5]{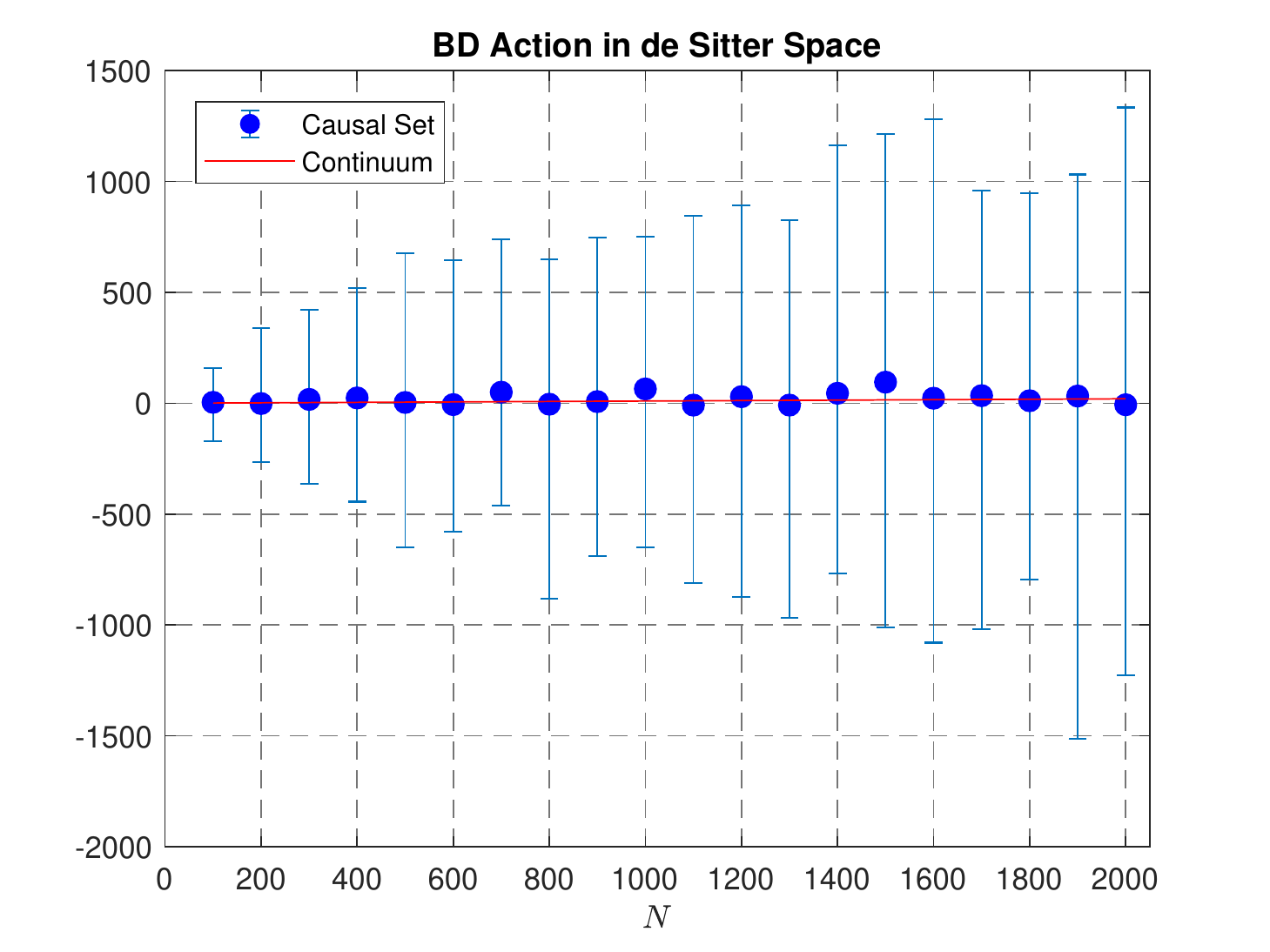}
\includegraphics[scale = .5]{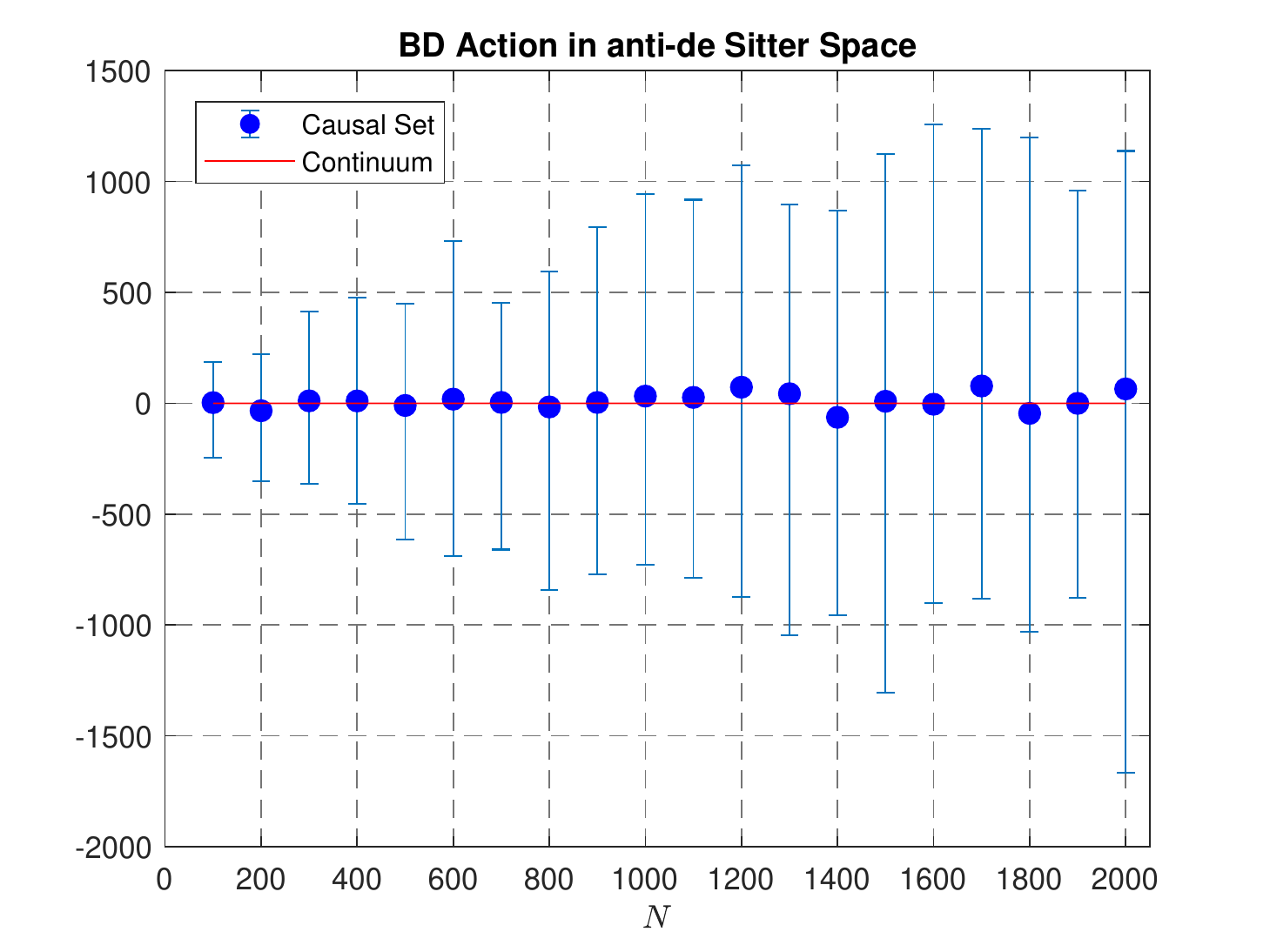}
\end{center}
\caption{Plots of the Benincasa-Dowker action vs.\ $N$ for Minkowski, de Sitter, and anti-de Sitter spaces. Everything else is the same as in the chain action case for comparison.}
\label{dact}
\end{figure}

\begin{figure}
\begin{center}
\includegraphics[scale = .5]{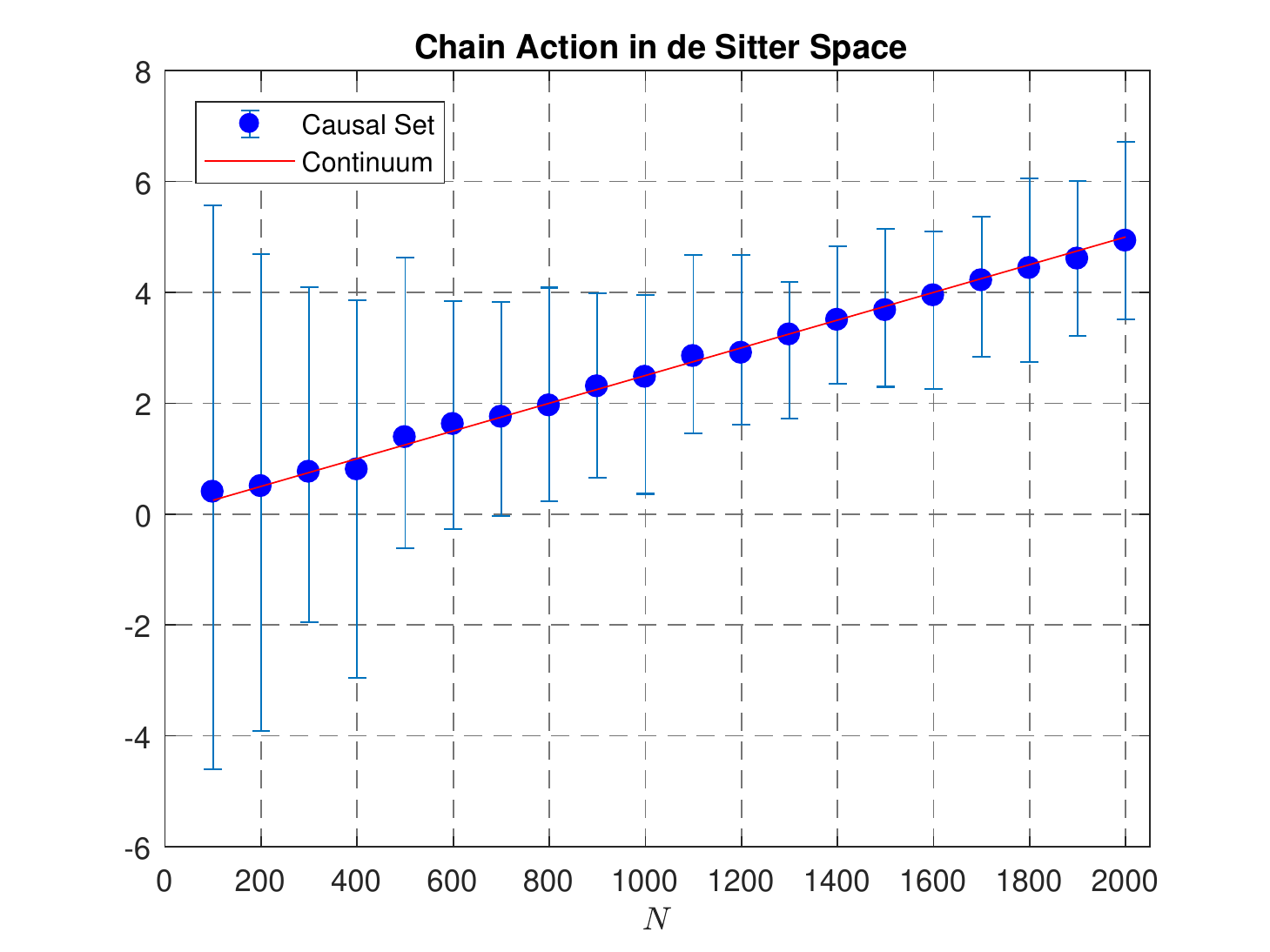}
\includegraphics[scale = .5]{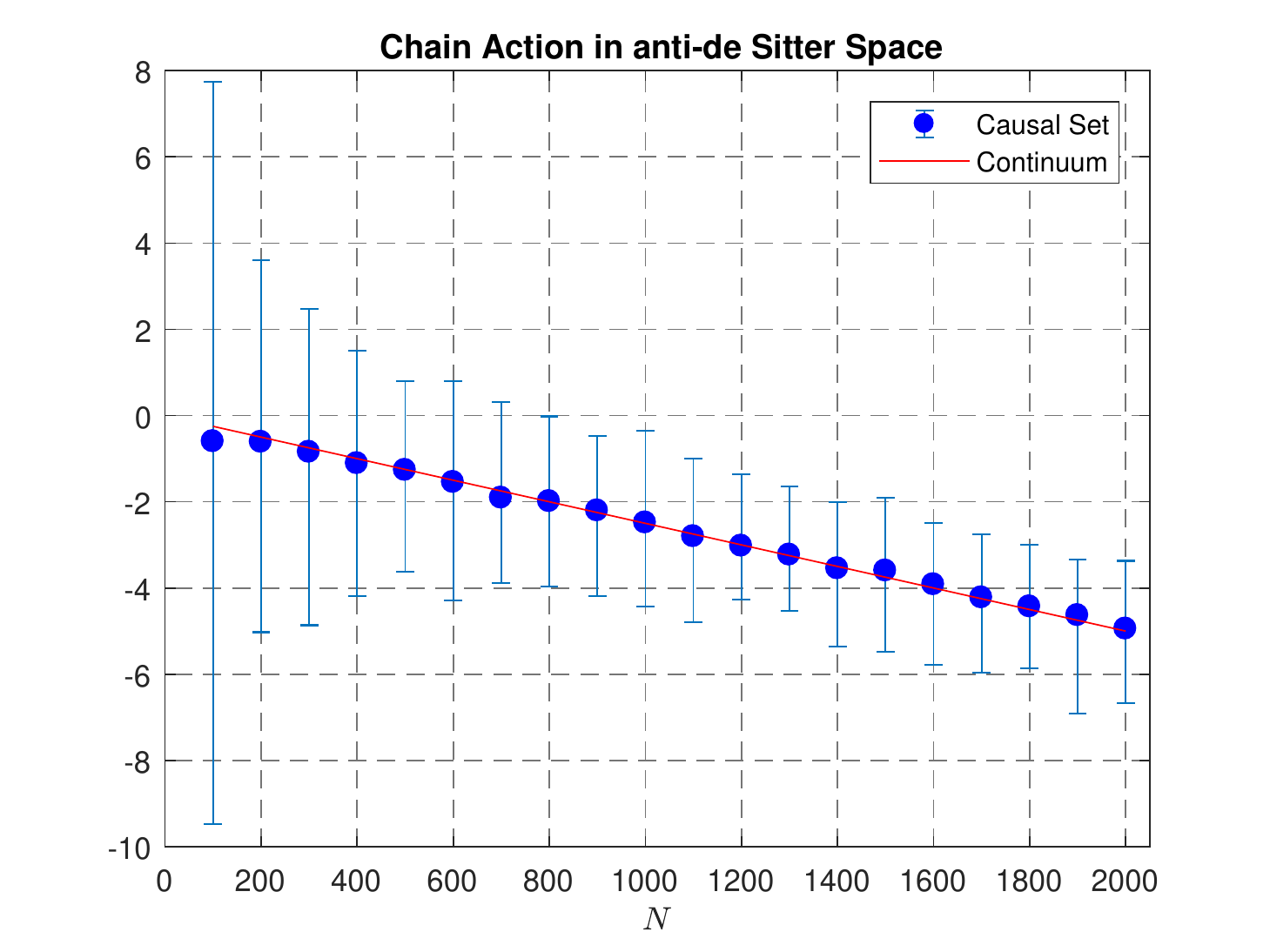}
\end{center}
\caption{Plots of the chain action for de Sitter and anti-de Sitter where $H=50$ and $\rho = 2\times 10^6$. In this case, $RL^2$ is smaller, and the assumption that $\mathcal{O}((\mathcal{R}L^2)^2)$ can be ignored is justified.}
\label{cact2}
\end{figure}

The results of this are in Fig.\ \ref{cact} and Fig.\ \ref{dact}. There's a lot to unpack here; both actions seem to mostly (we're not ignoring the deviation for large $N$ in de Sitter and anti-de Sitter; it will be addressed momentarily) on average match their continuum counterparts, but the chain action has some desirable properties the BD action lacks. The error bars for the chain action are relatively small, of order $1$, and more importantly, they shrink as $N$ increases. The BD action, however, has error bars of order $10^2-10^3$, and they increase as $N$ increases. The chain action does appear to have a flaw: for de Sitter and anti-de Sitter, it deviates from the continuum value, and this deviation increases with $N$. This is because in the derivation of the chain action, we neglected terms of $\mathcal{O}((\mathcal{R}L^2)^2)$, and for $N\approx 1000$, these terms start to become important for the values of $R$ and $\rho$ in this case. As the same assumption was made for the BD action, we suspect something similar happens, but the error bars are too large to notice it. As we can see in Fig.\ \ref{cact2}, if we choose a smaller value for the curvature, this complication doesn't occur.

\section{A Variational Principle}

We have now determined that both discrete actions match their continuum counterparts on average; what else can we do with an action? Ideally, we could set up a quantum version of the theory; however, before we do that, we should see if we can determine an analog of the field equations. Before we do this, we should explore what results we expect to get based on the continuum. In two dimensions, general relativity is trivial. The action is still the Einstein-Hilbert action,
\beq
S = \int R(x)\sqrt{-g(x)}\,\dd^2x \;,
\eeq
and the field equation is the Einstein equation which in the absence of matter reads
\beq
R_{\mu\nu}(x) - {\textstyle\frac12}\, g_{\mu\nu}(x)R(x) = 0 \;.
\eeq
This quantity is identically $0$ for any metric \cite{2dd,2dd2}; thus, we would expect all manifoldlike causal sets to obey our ``field equations."

\begin{figure}
\begin{center}
\includegraphics[scale = .5]{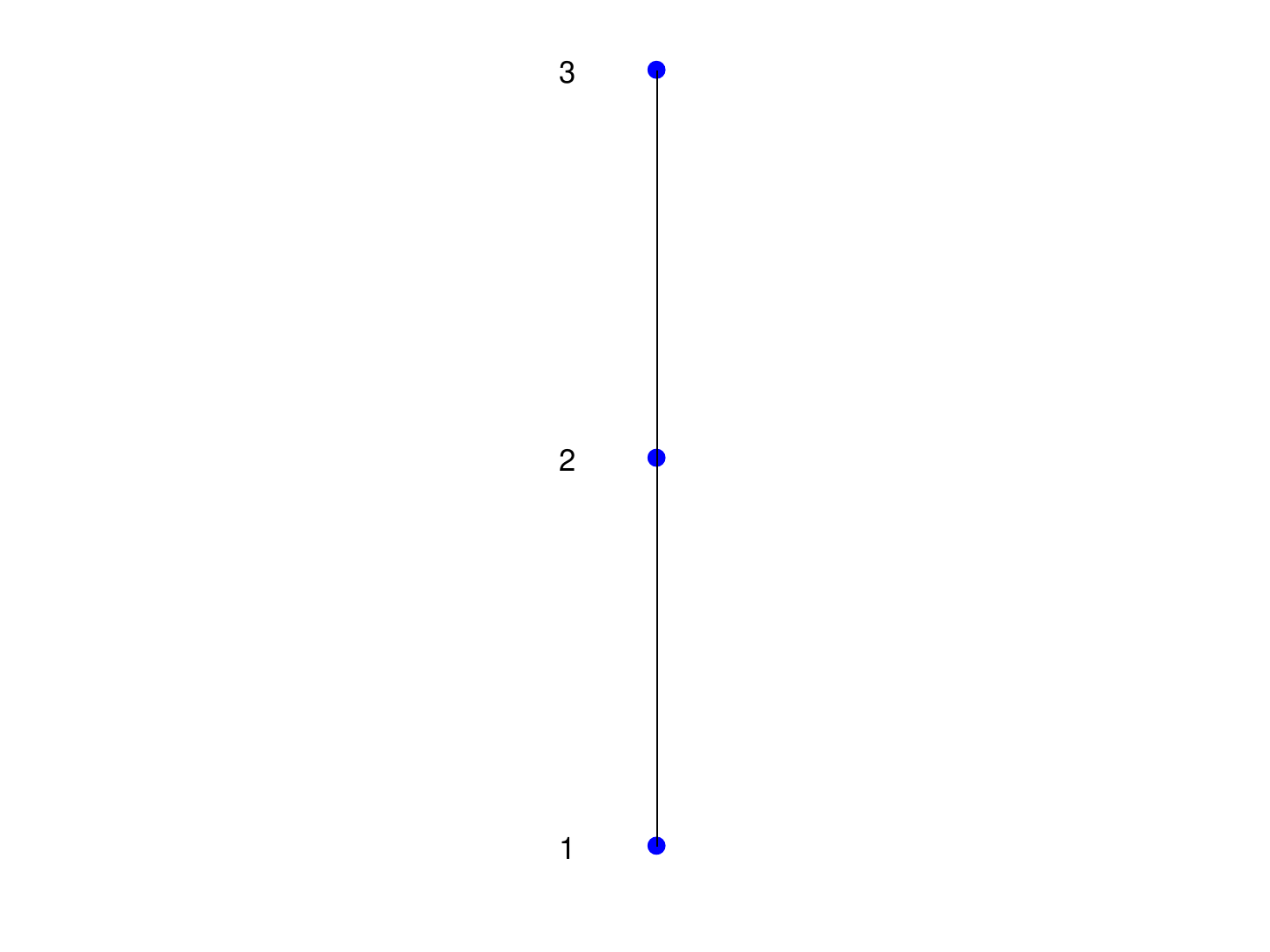}
\\
\includegraphics[scale = .5]{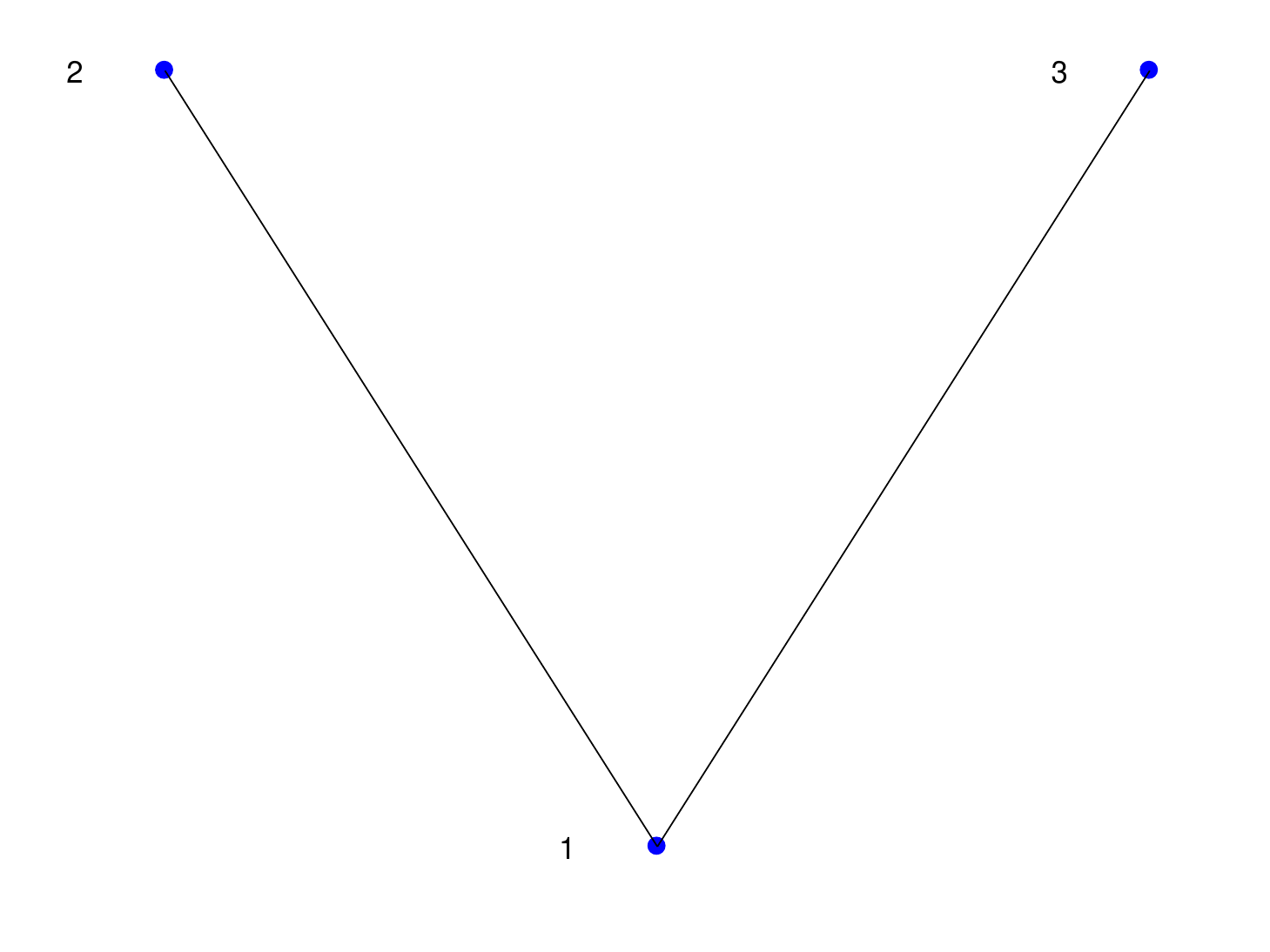}
\includegraphics[scale = .5]{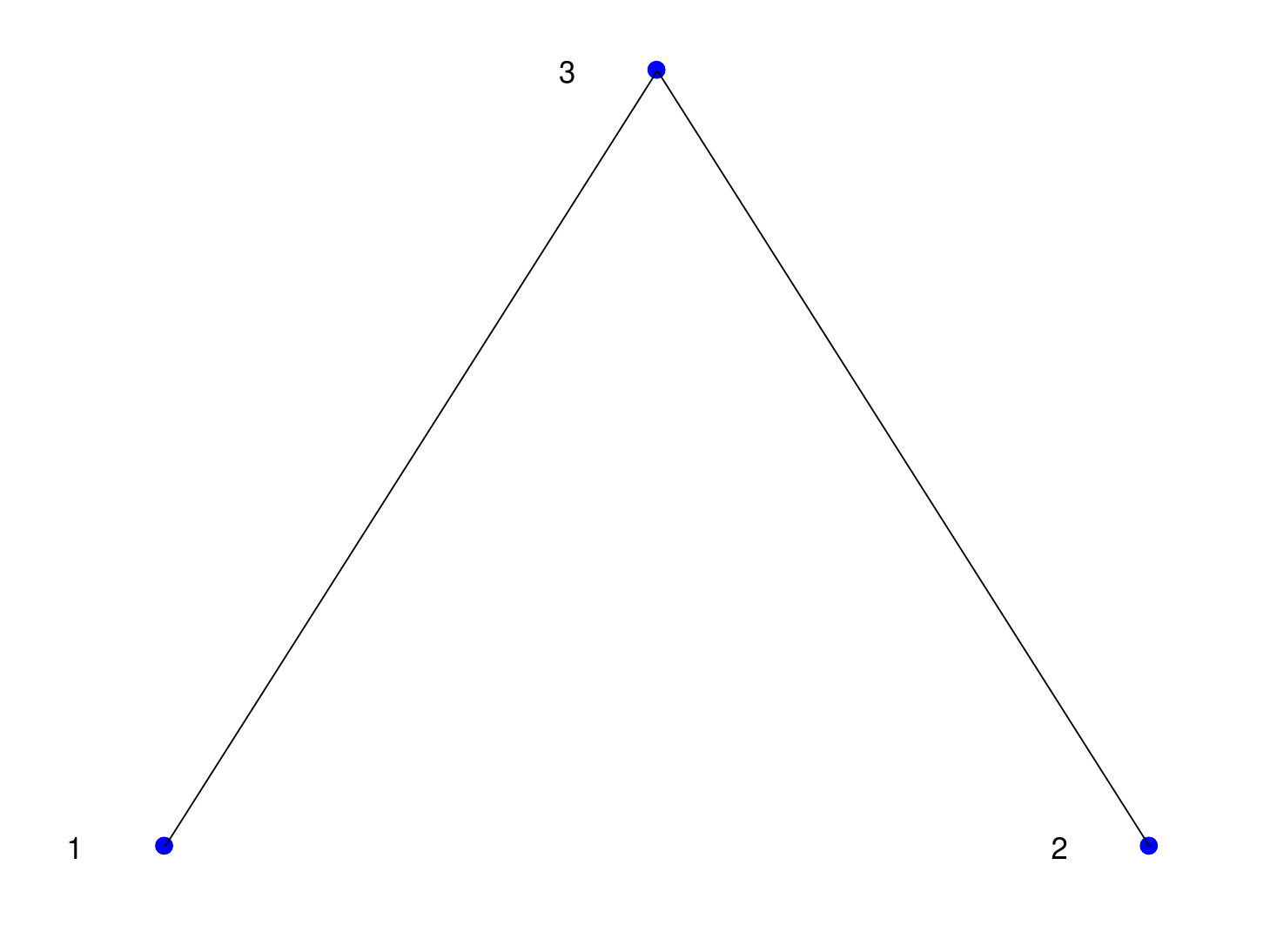}
\end{center}
\caption{A total chain (top). The other two causal sets can be generated from varying the total chain: $R_{2,3}\to0$ yields the image on the left, and $R_{1,2}\to 0$ yields the image on the right.}
\label{C}
\end{figure}

In a continuum theory the action is a functional of the field, $S[\phi(x)]$, and if we vary the field around a solution to the field equations, the resulting change of the action is zero. We can use this fact to determine the field equations. In causal set theory, we don't have this option; the variables of the theory are encoded into the relations matrix,\footnote{Or equivalently, the link matrix: $L_{i,j} = 1 \implies i\precc j$ and $0$ otherwise where $i\precc j \implies i\prec j$ and $R^2_{i,j} = 0$  In words, $i$ and $j$ are related with no points between them.} $R_{i,j}$. These variables are discrete and we can't vary them by infinitesimal amounts. We can instead make a discrete variation, under which we can't expect the change in the action $S$ to be zero and we can only look for those cases in which it is small. This requirement, however, is ambiguous and we need to define what we mean by a small change. As a necessary (but not sufficient) condition for $\Delta S$ to be small, we require that $|{\Delta S}/{S}| \ll 1$; as we will see, there may be additional requirements on its dependence on $N$. We will now explicitly state the process of variation: for (almost) each $R_{i,j}$ in turn, with $i$ between $1$ and $N$ and $j$ between $1$ and $N$ (we don't vary with respect to the minimal or maximal point to maintain an interval), we switch its value. If $R_{i,j} = 0$, we switch it to $1$ and vice versa. The only caveat is if $R_{i,j} = 1$, we don't consider $R_{j,i}$ a variable. If we switched it, $R_{j,i}\to 1$, then $i\prec j$ and $j\prec i$, but causal set theory expressly forbids closed timelike loops, so we disallow this possibility. Similarly, we don't vary $R_{i,i}$ which is always $0$. We're not quite finished; depending on the exact structure of the causal set, if we just switch one of the values of $R_{i,j}$, we could be left with something that isn't a causal set. We can illustrate this with an example. Consider some portion of a causal set with the relations matrix,
\beq
\mathbf{R} = 
\begin{pmatrix}
0 & 1 & 1\\
0 & 0 & 1\\
0 & 0 & 0
\end{pmatrix}.
\eeq
This represents a total chain displayed at the top of Fig.\ \ref{C}. There are three variables here: $R_{1,2}$, $R_{1,3}$, and $R_{2,3}$. If we vary $R_{2,3}\to 0$, we obtain the causal set on the bottom left of Fig. \ref{C}, and if we vary $R_{1,2}\to 0$, we obtain the causal set on the right. What if we vary $R_{1,3}\to 0$? In this case, we don't get a causal set; all causal sets must satisfy the transitive property: if $i\prec j$ and $j\prec k$, then $i\prec k$. In this case, we've violated this condition: $1\prec 2$ and $2\prec 3$, but $1\nprec 3$. If we want a causal set, we'll have to make at least one more change, either $R_{2,3}\to 0$ or $R_{1,2}\to 0$. As such we define our variational principle as follows: for all $R_{i,j}$ excluding those described above, switch its value, and then make the smallest possible number of changes to maintain transitivity. If this is degenerate (i.e. if there are multiple ways to make the order a causal set that require the same number of changes and that number is the minimum required,) choose randomly among the possibilities. 

How precisely can we do this? We could use brute force; once we decide to swap some value of one of the elements, $R_{i,j}$, we'll obtain a new relations matrix $\mathbf{R'}$. If this matrix does not represent a causal set, we could begin switching all of the elements one at a time, two at a time, three at a time, etc.\ until we obtain a valid relations matrix. We then find all of the valid causal sets with this number of changes and randomly select one. This is a perfectly legitimate method; however, as one might imagine, if we wrote a computer program that did this for a $1000$ element causal set, we'd all be dead before it finished. Instead, we should find a smarter way to do this. There are two cases, $R_{i,j}\to 0$ and $R_{i,j}\to 1$.
\begin{enumerate}
\item $R_{i,j}\to 0$\\
In this case, we have two elements of the causal set that are related, and we wish to delete the relation and change any others necessary to preserve transitivity. To do this, we only need to concern ourselves with the elements between $i$ and $j$; we use $\prec$ to specify the partial order of the original causal set and $\tilde{\prec}$ for the new causal set we want to build. We must split the elements between $i$ and $j$ into two categories: those related to $i$ and not $j$, $K = \{k\mid i\tprec k, k\tnprec j\}$ and those related to $j$ but not $i$, $L = \{l\mid i\tnprec l, l\tprec j\}$ where all of the elements of both sets are between $i$ and $j$ in the original causal set. How do we choose which elements belong to which set? To achieve the fewest number of changes, either $L = \varnothing$ and $K = \{k\mid i\prec k \prec j\}$ or vice versa. In words, we break the relations between all of the elements between $i$ and $j$ from either $i$ or $j$. There's a simple argument that this is the smallest number of changes necessary: no matter how we construct $L$ and $K$, the number of changes necessary is $\#(L) +\#(K) + \text{mixed terms}$ where the mixed terms are the number of relations from $l\in L$ to $k \in K$. $\#(K)+\#(L)=m$ is always the same; it's the number of elements between $i$ and $j$ in the original causal set, $m$. Thus, to minimize the number of changes, we need to minimize the number of mixed terms; to do this, we let either $K$ or $L$ be the set of $m$ elements between $i$ and $j$.
\item $R_{i,j}\to 1$\\
In this case, the only elements that matter are those to the past of $i$, $P(i) = \{k\mid k\prec i\}$ and those to the future of $j$, $F(j) = \{k\mid j\prec k\}$. For a similar argument to the other case, the way to generate the new causal set with the fewest number of changes is to either eliminate all of the relations between the elements of $P$ from $i$ and add relations between $i$ and the elements of $F$ or eliminate all relations between $j$ and the elements of $F$ and add relations between the elements of $P$ and $j.$\\
\end{enumerate}

This method can cause a relatively large number of relations to be changed, so an obvious question is, does this constitute a small change? From the point of view of the continuum, a small change could be achieved by moving the location of an element of the causal set by some small amount and recalculating its causal relationships with the other elements; however, we would like our formulation to only refer to causal set quantities so it can be applied to a generic causal set and not just the manifoldlike ones. With only the variables of the causal set, our formulation causes the least number of changes to the original configuration without expressly forbidding variations with respect to variables which cause a ``large'' change however one would like to define that. We find that to be a unsatisfying, but if there is a more appropriate definition of a small change which is less impactful than this, it at least provides an upper bound on $\Delta S$.

\subsection{Simulations}

We use the above prescription on $100$ causal sets of each size in different spacetimes for both actions to find plots of the average value of $|{\Delta S}/{S}|$ as a function of $N$. There is a different $\Delta S$ for each $R_{i,j}$ we vary; as such we choose the one with the largest $|\Delta S|$ to make the plots (if the largest $|\Delta S/S|$ is small, the rest must be as well.)
\begin{figure}
\begin{center}
\includegraphics[scale = .5]{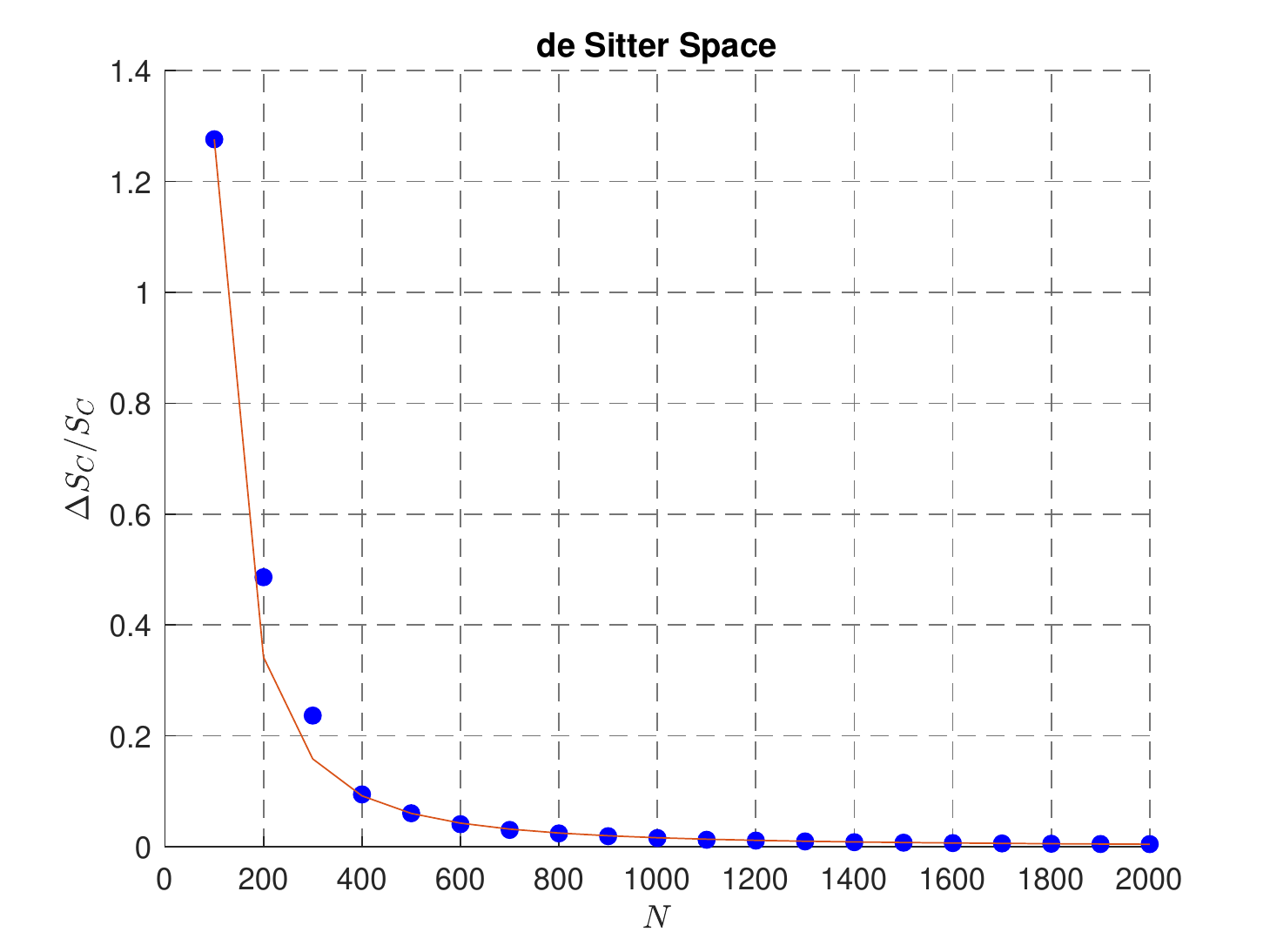}
\includegraphics[scale = .5]{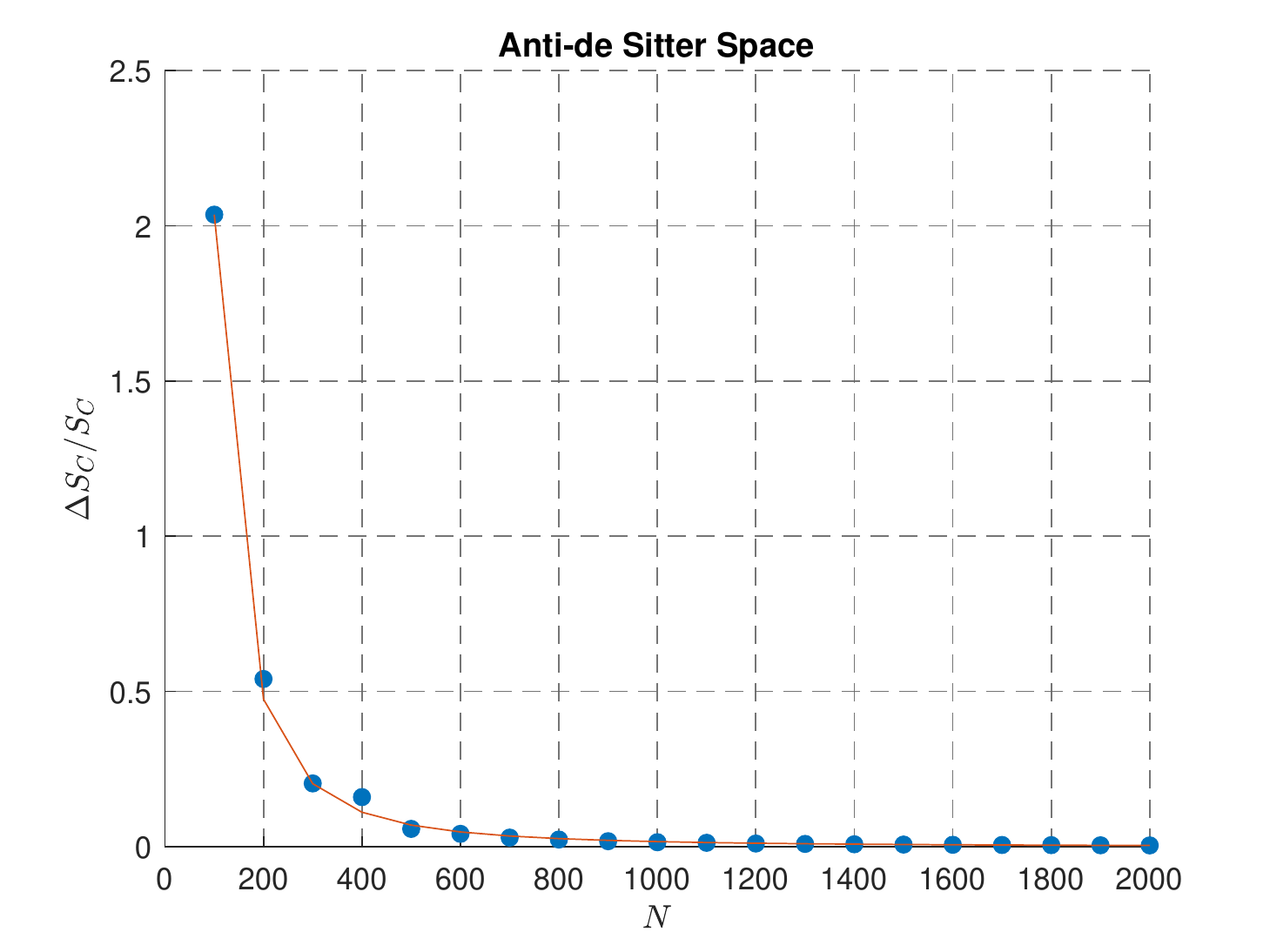}

\includegraphics[scale = .5]{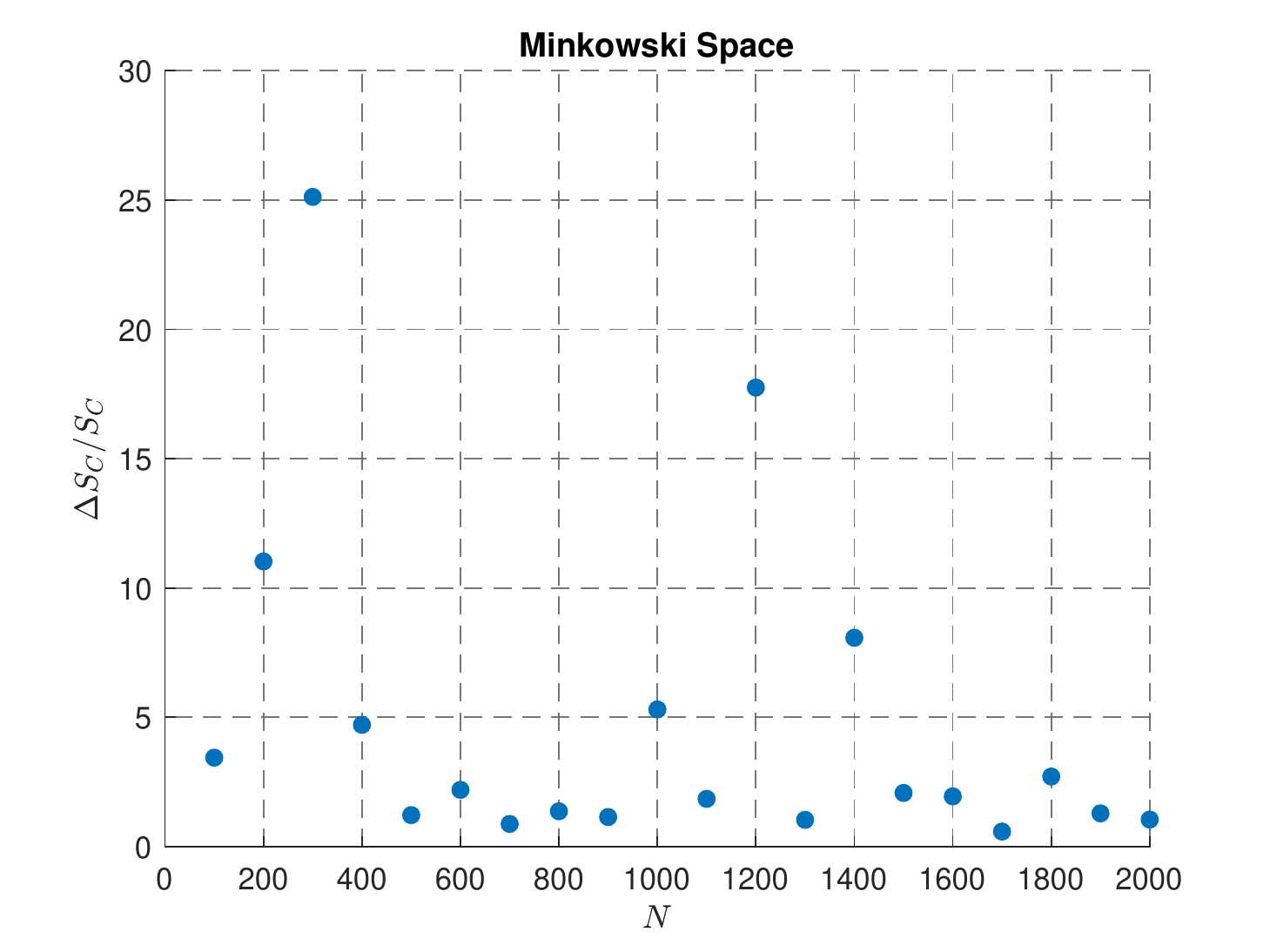}
\end{center}
\caption{A plot of $|\Delta S\_{C}/S\_{C}|$ vs.\ $N$ for de Sitter with $H = 100, \rho = 2\times 10^6$ (top left), anti-de Sitter with $H = 100, \rho = 2\times 10^6$ (top right), and Minkowski space (bottom). The orange lines are attempted fits with $N^{-1.9}$ and $N^{-2.1}$ dependences, respectively.}
\label{cSoS}
\end{figure}
\begin{figure}
\begin{center}
\includegraphics[scale = .5]{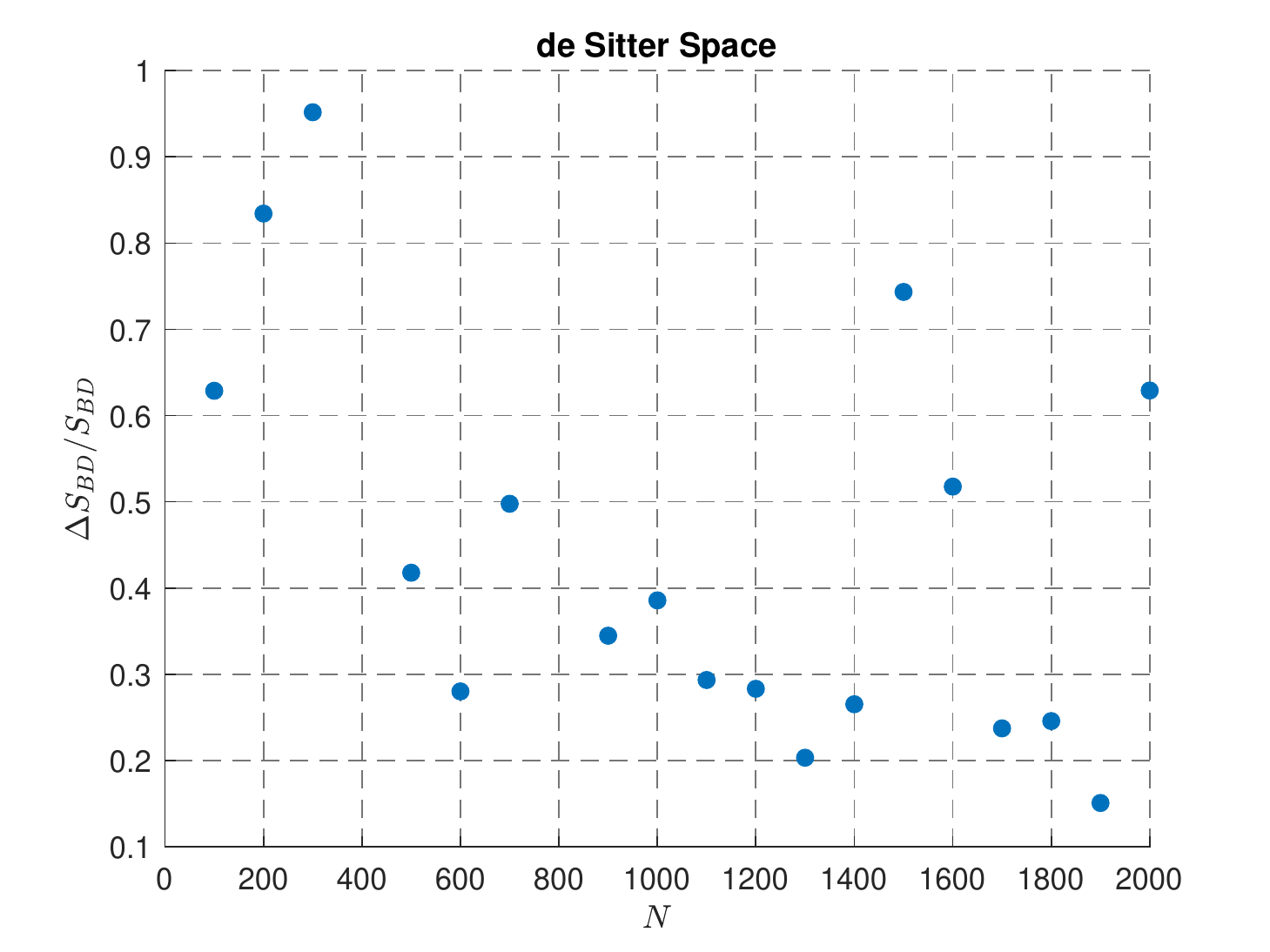}
\includegraphics[scale = .5]{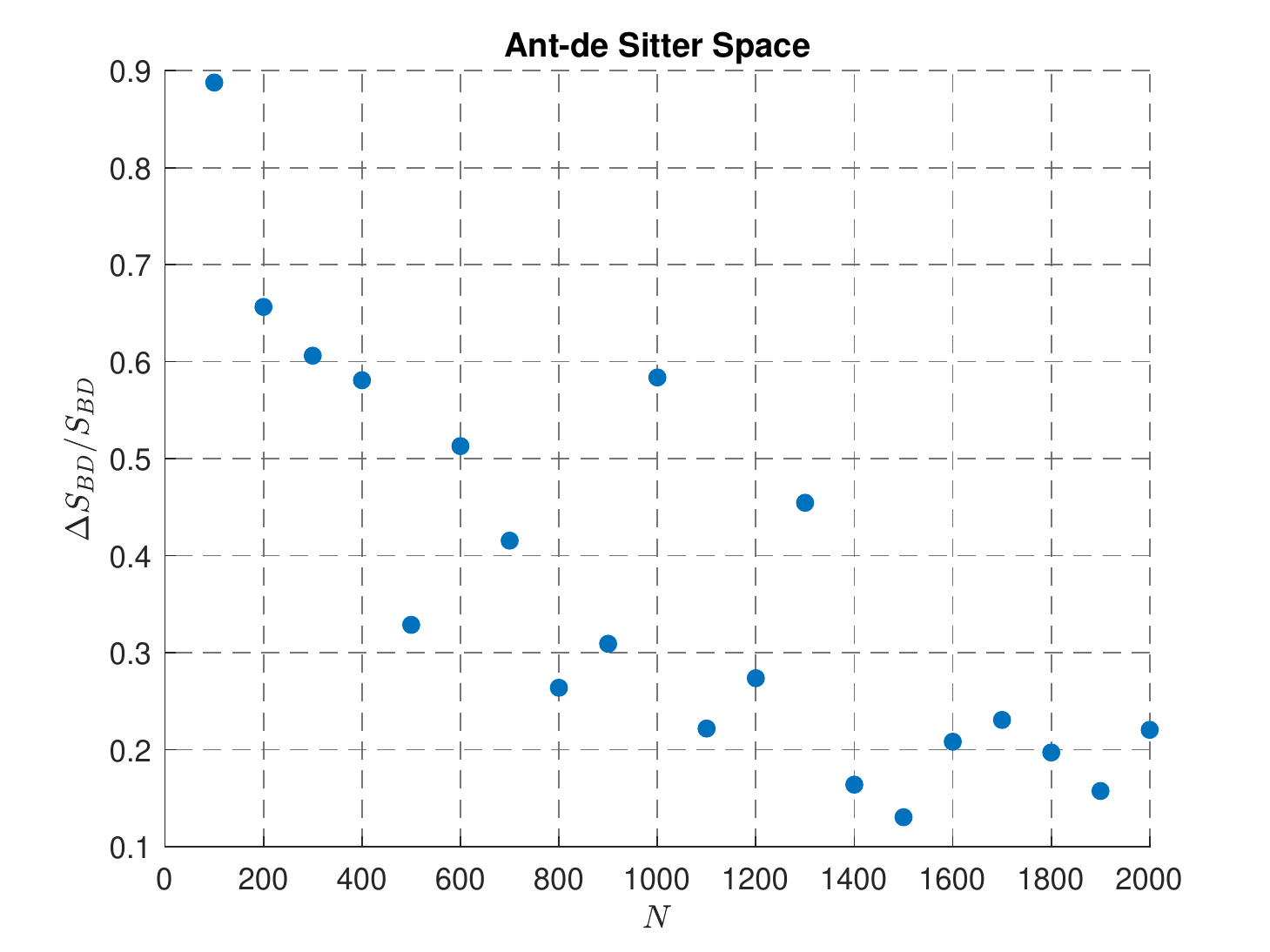}

\includegraphics[scale = .5]{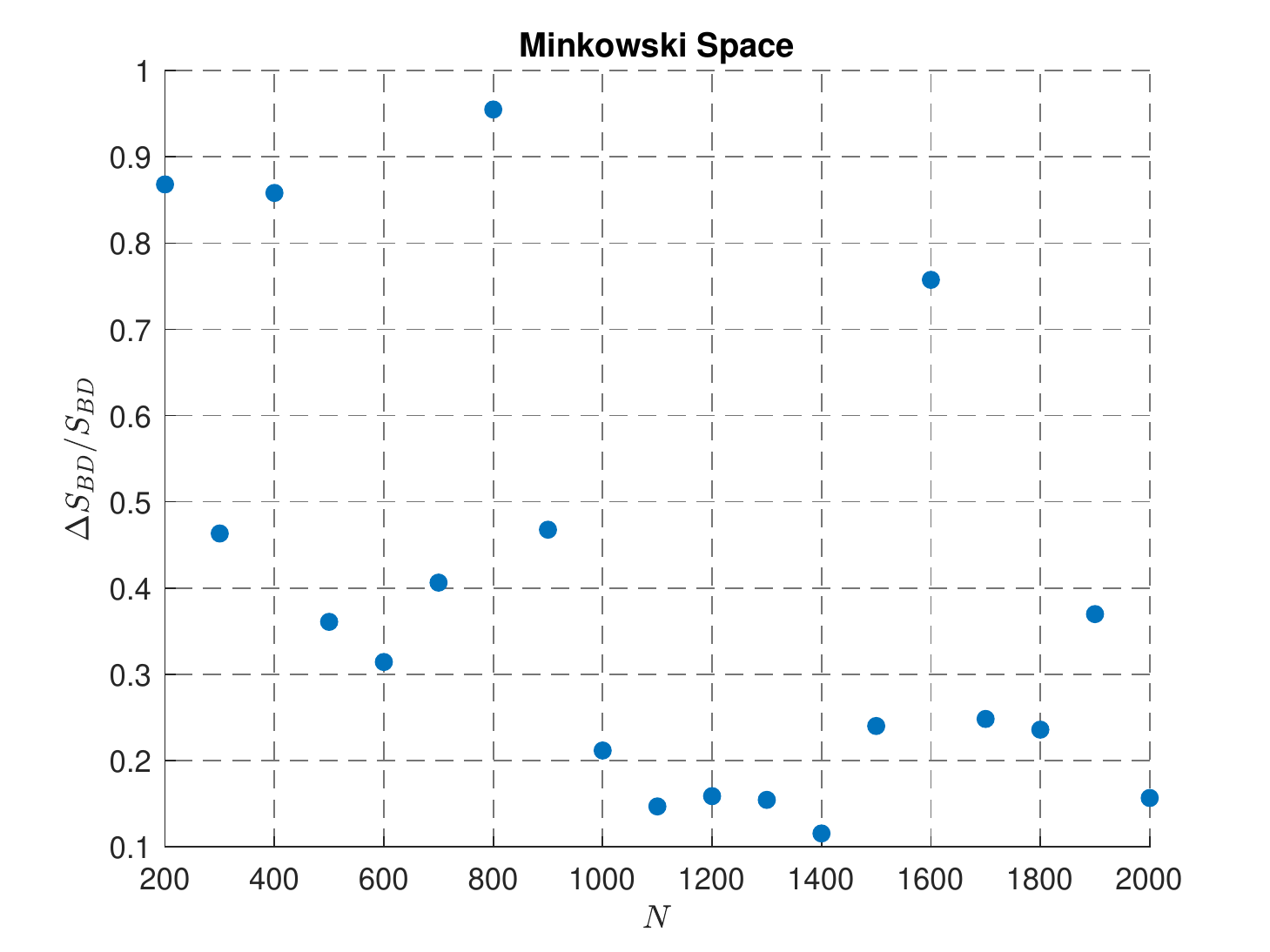}
\end{center}
\caption{A plot of $|\Delta S\_{BD}/S\_{BD}|$ vs.\ $N$ for de Sitter with $H = 100, \rho = 2\times 10^6$ (top left), anti-de Sitter with $H = 100, \rho = 2\times 10^6$ (top right), and Minkowski space (bottom).}
\label{cSoS}
\end{figure}

Exploring the chain action first, we can see that for de Sitter and anti-de Sitter, $|\Delta S\_{C}/S\_{C}|$ falls off very quickly. In de Sitter space this curve seems to follow a $1/N^{1.9}$ behavior and in anti-de Sitter space a $1/N^{2.1}$ behavior. This leads us to believe that in general $|\Delta S\_{C}/S\_{C}| \sim (1/N)^{f(R)}$ plus higher-order terms where $f(R)$ is some function of $R$ likely centered around $2$, but this needs more study. Minkowski space, however, identifies a problem with this scheme: what if the action is close to $0$? In this case, the quotient can be arbitrarily large;\footnote{For a continuum theory, this wouldn't really be a problem. No matter how small $S$ is, $\Delta S$ can just be much smaller, but in this case, $\Delta S$ has a minimum value.} more work needs to be done to explore this case.

\begin{figure}
\begin{center}
\includegraphics[scale = .5]{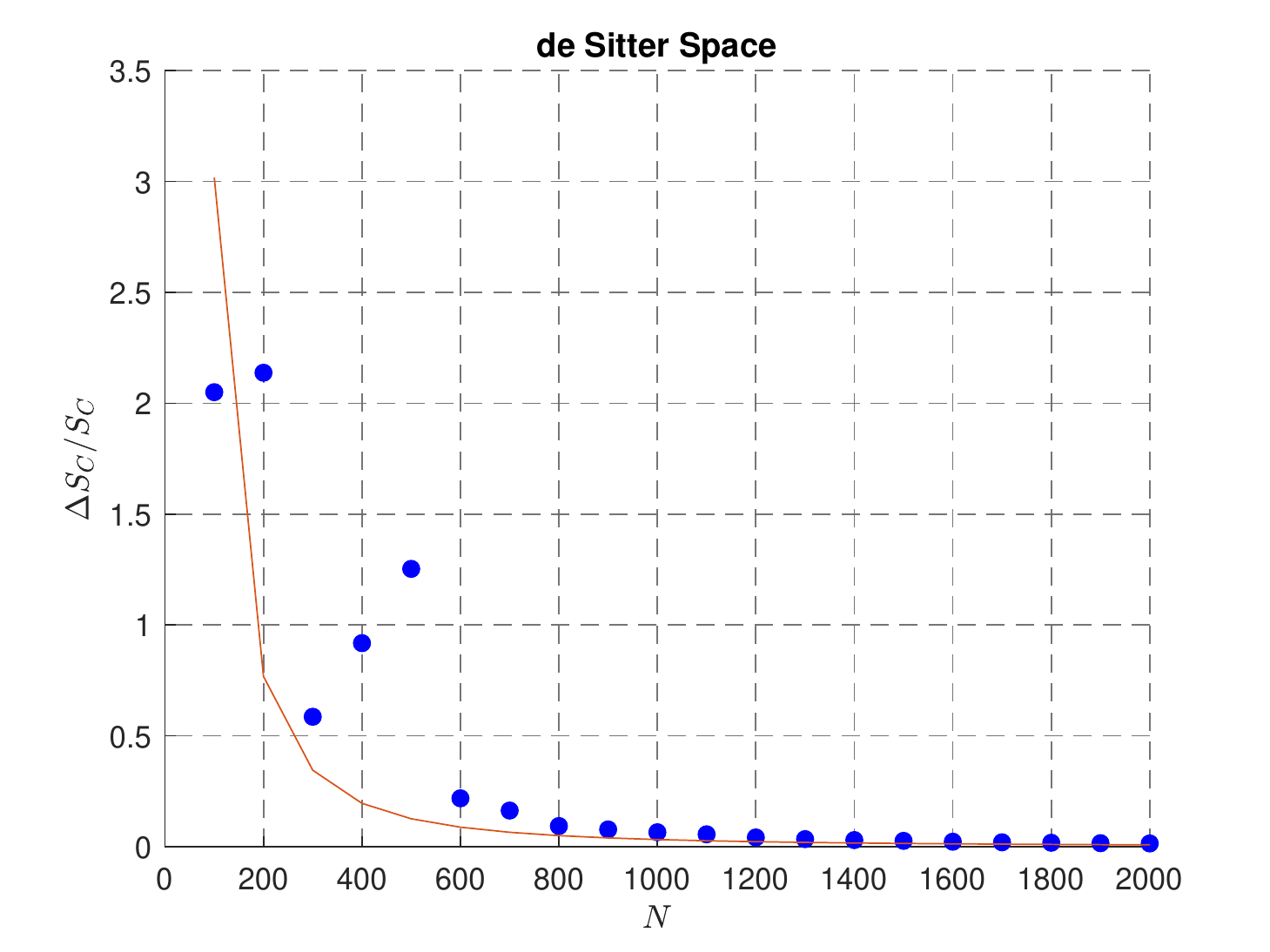}
\end{center}
\caption{A plot of $|\Delta S\_{C}/S\_{C}|$ in de Sitter for $H = 50$ and $\rho = 2 \times 10^6$. For this case, we can see that for low values of $N$, the plot is skewed we suspect from some sprinklings where $S$ is very close to $0$, but the later points are still fit very well by $1/N^{1.9}.$}
\label{badsc}
\end{figure}

\begin{figure}
\begin{center}
\includegraphics[scale = .5]{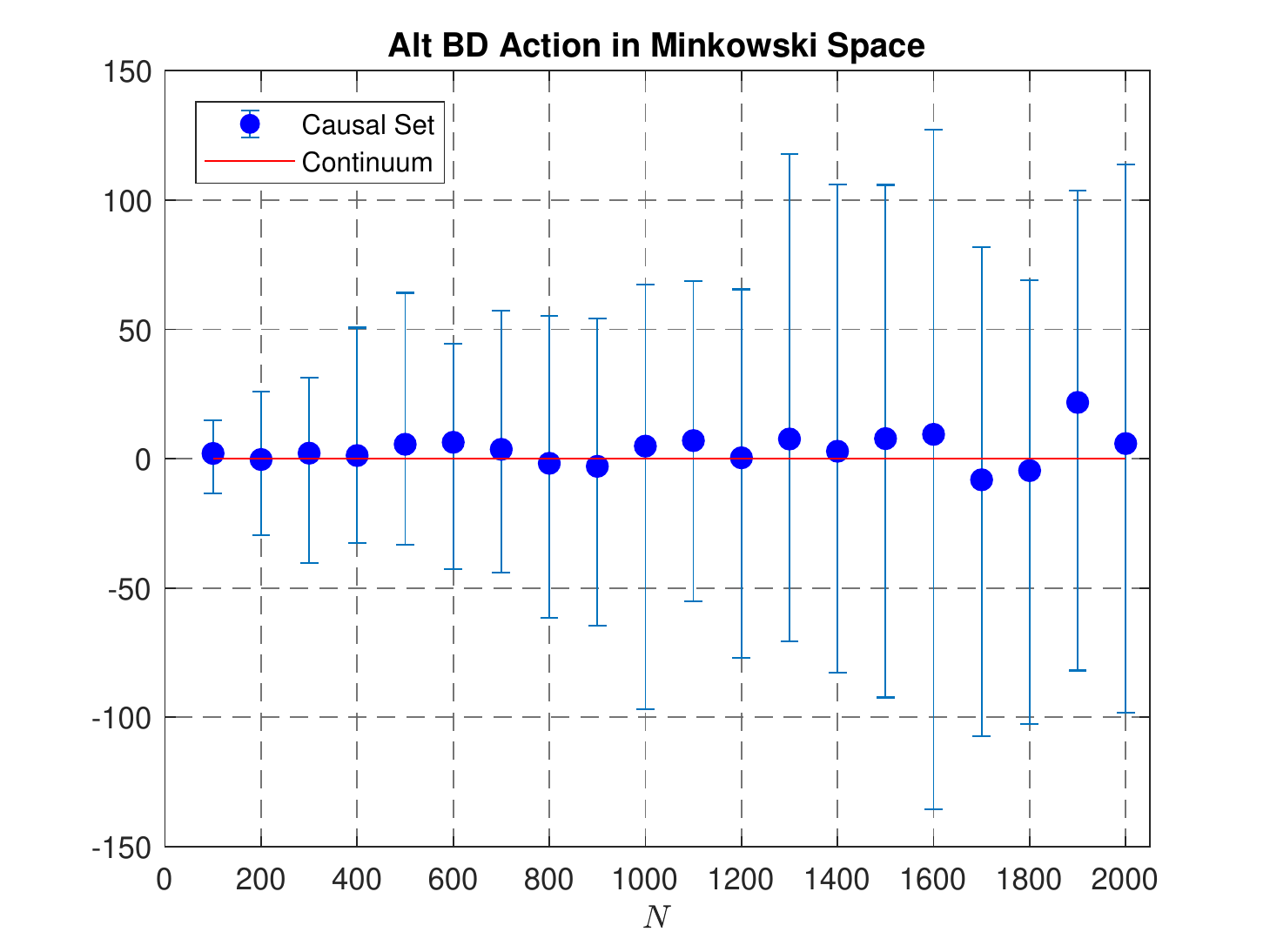}
\end{center}
\caption{A plot of the alternative version of the BD action vs. $N$ for $\epsilon = .2$ in Minkowski space averaged over $100$ causal sets of each size.}
\label{altbd}
\end{figure}

The BD action has the same problem for all spacetimes; although there is a general downward trend, there is no obvious behavior with $N$. We suspect the plots are skewed by values of the action which are very close to $0$ which don't occur for these values of $H$ and $\rho$ in the chain action (except at very low values of $N$); however, for other values of $H$ and $\rho$ the chain action can still have this problem as seen in Fig.\ \ref{badsc}. It's worth noting that there's a form of the BD action that depends on an additional parameter $\epsilon$, given by \cite{Glaser}
\beq
S_{\text{BD},\epsilon} = 4\epsilon\bigg(N-2\epsilon\sum_{n=1}^{N-1}N_nf(n,\epsilon)\bigg),
\eeq
where 
\beq
f(n,\epsilon) = (1-\epsilon)^{n-1}\left(1-\frac{2\epsilon (n-1)}{1-\epsilon}+\frac{\epsilon^2 (n-1)(n-2)}{2(1-\epsilon)^2}\right),
\eeq
and the parameter $\epsilon$ defines an intermediate length scale $\ell/\sqrt\epsilon$ between $\ell$ and the linear scale of the Alexandrov set. As seen in Fig.\ \ref{altbd}, this form does lower the error bars, but qualitatively its features are the same. 


\begin{figure}
\begin{center}
\includegraphics[scale = .5]{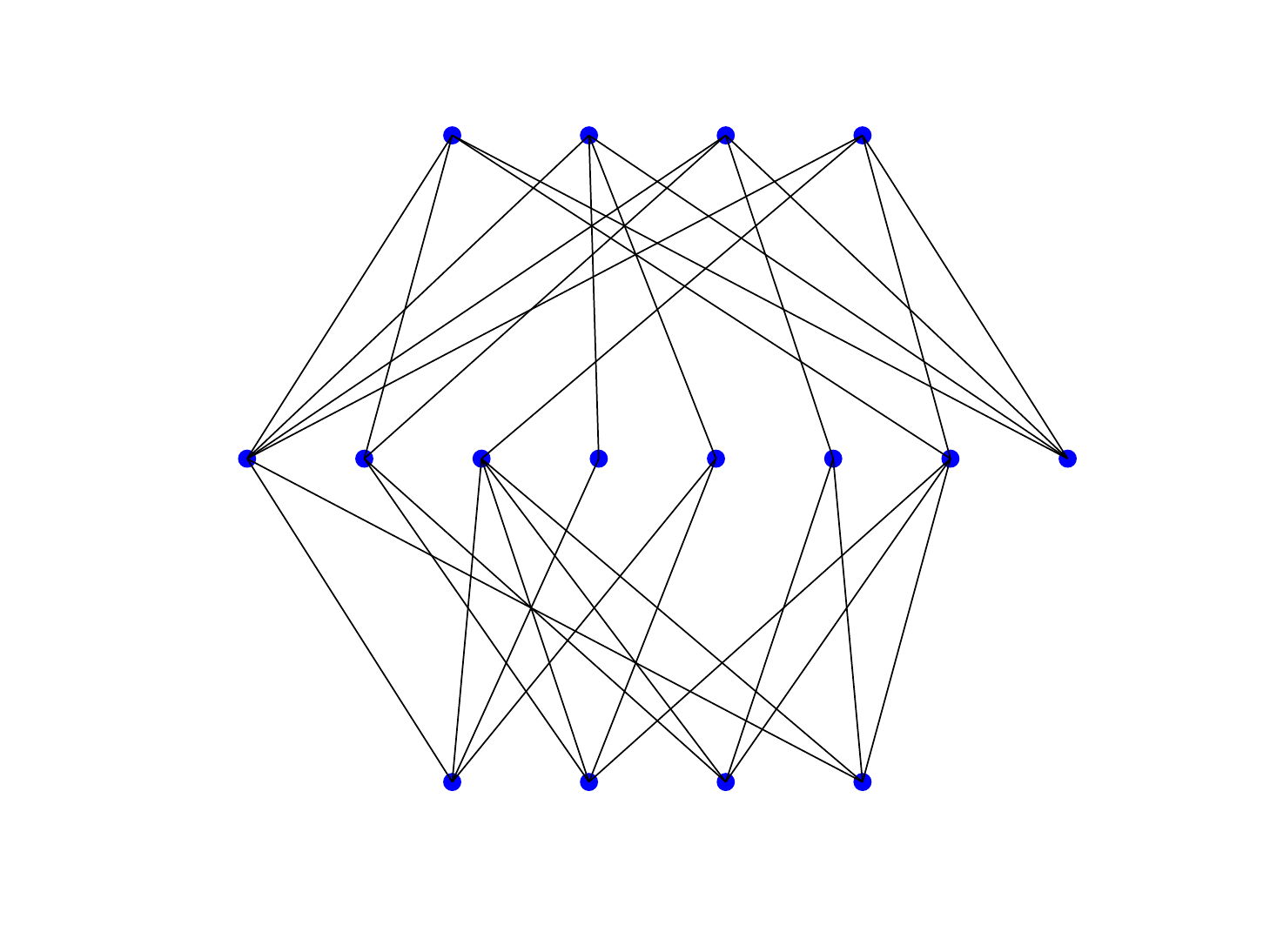}
\includegraphics[scale = .5]{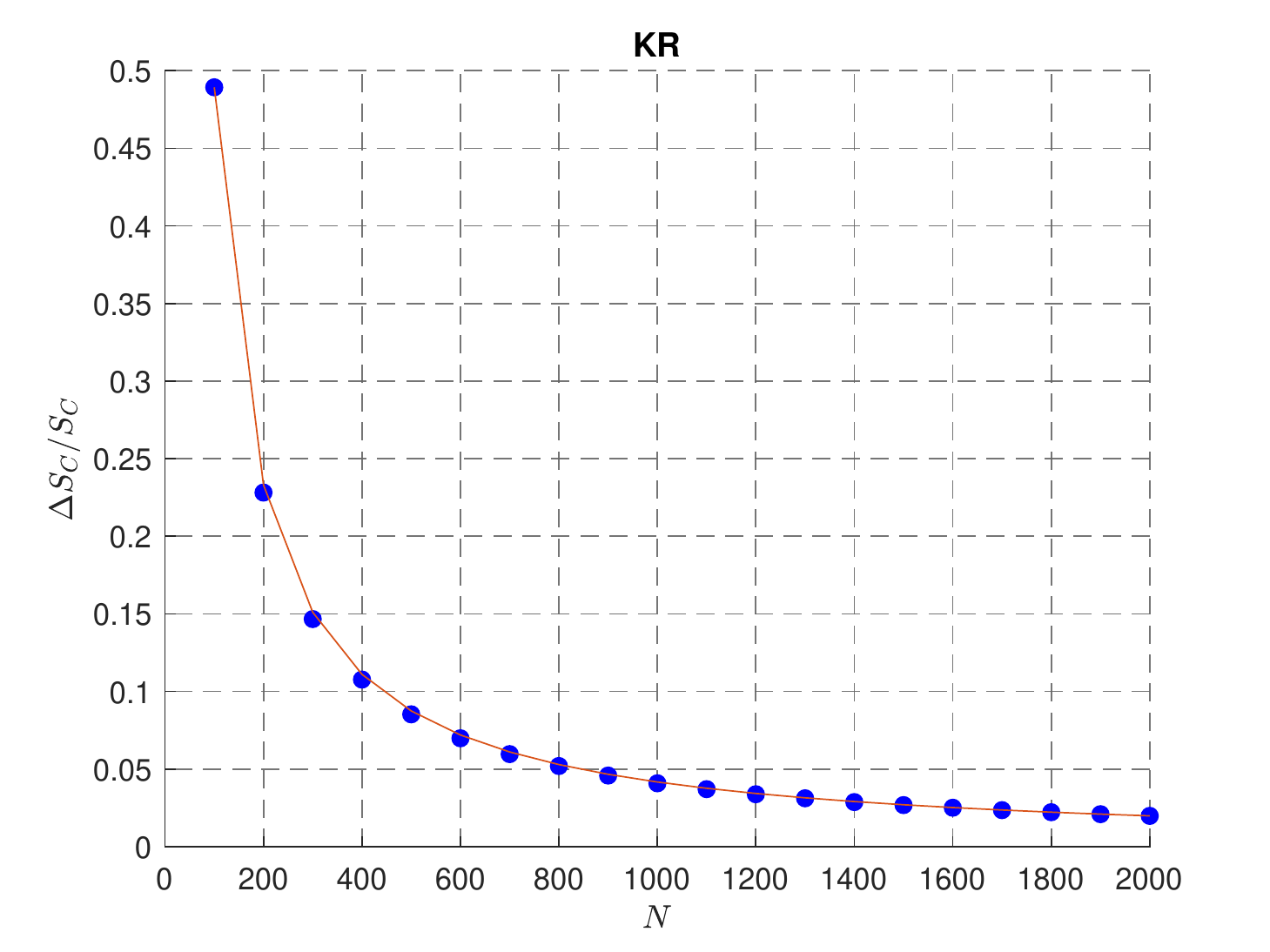}
\end{center}
\caption{An example of a $16$-point KR causal set with the links drawn (left). A plot of $|\Delta S\_{C}/S\_{C}|$ vs.\ $N$ averaged over $100$ KR causal sets (right).}
\label{KlR}
\end{figure}

At least for the chain action, it seems that in cases where $S$ is far from $0$, on average $\Delta S/S\to 0$ for large $N$, but to be sure that from this we can conclude that these sprinklings satisfy the discrete field equation, we should explore an example of causal sets which shouldn't satisfy this equation. Because the Einstein tensor is identically $0$ for all metrics in two dimensions, that leaves non-manifoldlike causal sets. The most prominent example of these are the Kleitman-Rothschild (KR) causal sets; for large $N$ these represent the vast majority of causal sets (as $N\to \infty$, the fraction of causal sets which are of this type tends to $1$ \cite{KR}). These KR causal sets have 3 layers, with $N/4$ elements in the top and bottom layers and the remaining $N/2$ elements in the middle layer; futhermore, each point in the top and bottom layers is related to half of the points in the middle. A $16$-point example is shown on the left of Fig.\ \ref{KlR}; to define the chain action here, we add an overall maximal and minimal point. We can then treat these the same way as the other causal sets: by simulating $100$ causal sets of each size and averaging $|\Delta S/S|$. The results are shown on the right of Fig.\ \ref{KlR}. The fit in this case goes like $1/N^{1.07}$; this still approaches $0$ in the $N\to\infty$ limit, but for any large but finite value of $N$, all three of the manifoldlike plots where $S\neq 0$ are much smaller. In the regime where $R$ is close to constant over the Alexandrov set, all manifolds should approximate de Sitter or anti-de Sitter depending on the sign of the scalar curvature, so we expect this to hold in general. 

\section{Conclusion}

We've shown that both the Benincasa-Dowker and chain actions on average approximate the continuum Einstein-Hilbert action at least in the regime in which we're interested. We've also proposed a discrete analog of a variational principle which in the case of the chain action seems to distinguish $S\neq 0$ manifoldlike causal sets from the most common type of nonmanifoldlike causal sets. Unfortunately the large error bars for the BD action prevent this from happening at least for scalar curvatures of the order we've tested.\footnote{It seems likely we could make this work for the BD action if we used very large scalar curvatures and very large densities, but since the errorbars grow with $N$, eventually the same problem would emerge.} The BD action does have two major advantages over the chain action, however. Thus far, we've only considered the Einstein-Hilbert action though it's well known this is not the whole story; in cases where there are boundaries, one must include the Gibbons-Hawking-York term \cite{GHY}, and there's some evidence that the BD action already includes this term \cite{BT}. The chain action does not; however, it's possible that including a suitable boundary term would fix the $S=0$ case. This is a task for future work. We would also like to extend this to higher dimensions where the Einstein equation isn't trivial; however, in higher dimensions, the only vacuum solutions have $S = 0$, so the boundary problem must be addressed first.

Furthermore, although neither the BD action nor the chain action are local, the chain action is in some sense less local. In the continuum, one can define a Lagrangian density (the argument which when integrated over produces the action) as a particular combination of the metric and its derivatives at a single point. In the case of causal sets, this is of course not possible, so some amount of nonlocality is necessary. As an example, consider a causal set embedded in a very large portion of Minkowski space. Each point $i$ far from the past boundary will have a very large number of related points to its past and thus also a very large number of intervals of each size. However, in the BD action only intervals up to size three are considered; this is the causal set equivalent of using not only a point $x_i$, but also the union of all Alexandrov sets with some constant height that terminate at $x_i$ to define the Lagrangian density. Therefore, the Lagrangian density at $x_i$ depends on points an arbitrary coordinate distance away though they are bounded by some hyperboloid a (small) proper time away, thus maintaining some form of locality. For the chain action though, there will be chains of length three both an arbitrary coordinate distance and proper time away. It's unclear what effect this greater nonlocality of the chain action has. One possible way to address it is to only consider maximal chains (in which the interval between any two consecutive points is empty), since short maximal chains also extend only a short proper time away. This is another task for future work.

\end{document}